\documentclass[apj]{emulateapj}
\usepackage{natbib}
\usepackage{float}
\pdfoutput=1

\begin{document}

\title{Auroral radio emission from late L and T dwarfs: 
        \\ A new constraint on dynamo theory in the substellar regime}


\author{Melodie M. Kao\altaffilmark{1,2},
        Gregg Hallinan\altaffilmark{2},
        J. Sebastian Pineda\altaffilmark{2},
        Ivanna Escala\altaffilmark{3},
        Adam Burgasser\altaffilmark{3}, 
        Stephen Bourke\altaffilmark{2} \& 
        David Stevenson\altaffilmark{4}}

\altaffiltext{1}{Corresponding author: \email{mkao@astro.caltech.edu} }
\altaffiltext{2}{California Institute of Technology, Department of Astronomy, 1200 E California Blvd, MC 249-17, Pasadena, CA 91125, USA}
\altaffiltext{3}{University of California San Diego, Center for Astrophysics and Space Sciences, 9500 Gilman Drive, MC 0424, La Jolla, CA 92093, USA}
\altaffiltext{4}{California Institute of Technology, Division of Geological \& Planetary Sciences, 1200 E California Blvd, MC 150-21, Pasadena, CA 91125, USA}



\begin{abstract}
We have observed 6 late-L and T dwarfs with the Karl G. Jansky Very Large Array (VLA) to investigate the presence of highly circularly polarized radio emission, associated with large-scale auroral currents. Previous surveys encompassing $\sim$60 L6 or later targets in this spectral range have yielded only one detection. Our sample includes the previously detected T6.5 dwarf 2MASS~10475385+2124234 as well as 5 new targets selected for the presence of H$\alpha$ emission or optical/infrared photometric variability, which are possible manifestations of auroral activity. We detect 2MASS~10475385+2124234, as well as 4 of the 5 targets in our biased sample, including the strong IR variable SIMP~J01365662+0933473 and bright H$\alpha$ emitter 2MASS~12373919+6526148, reinforcing the possibility that activity at these disparate wavelengths is related. The radio emission frequency corresponds to a precise determination of the lower-bound magnetic field strength near the surface of each dwarf and this new sample provides robust constraints on dynamo theory in the low mass brown dwarf regime. Magnetic fields $\gtrsim2.5$~kG are confirmed for 5/6 targets.  Our results provide tentative evidence that the dynamo operating in this mass regime may be inconsistent with predicted values from a recently proposed model.  Further observations at higher radio frequencies are essential for verifying this assertion. 
\end{abstract}

\keywords{brown dwarfs --- planets and satellites: aurorae ---
planets and satellites: magnetic fields --- radio continuum: stars --- stars: individual (2MASS~10430758+2225236, 2MASS~12373919+6526148, SDSS~04234858-0414035, SIMP~J01365662+0933473) --- stars: magnetic fields}



\section{Introduction}\label{sec.intro}
An important outstanding problem in dynamo theory is understanding how magnetic fields are generated and sustained in fully convective stellar objects.  Prevailing dynamo models for dwarf stars with an inner radiative zone and an outer convective envelope, like the Sun, are accepted to rely on the shearing at the interface between these two layers, where differential rotation is strongest \citep{parker1975}.  Beyond spectral type $\sim$M4, stars are fully convective and no longer possess the internal structures necessary to sustain such dynamos \citep{chabrierBaraffe1997}. However, flaring M-dwarfs are characterized by kilogauss fields covering much of the stellar disk \citep{saar1994, johnskrullValenti1996}, and the fraction of M, L and T dwarfs that exhibit strong and persistent H$\alpha$ emission, a magnetic activity tracer, rises through late-M dwarfs and peaks at $\sim$90\% for L0 dwarfs before declining to $\sim$50\% for L5 dwarfs \citep{gizis2000, west2004, west2008, schmidt2015}. Clearly, an alternative dynamo operates in low mass, fully convective stars. A number of models for possible dynamo mechanisms in this regime have been proposed \citep{chabrierKuker2006, dobler2006, browning2008, christensen2009, morin2011, gastine2013}, but constraining data on magnetic field strengths and topologies across a wide range of mass, age, rotation and temperature are sorely lacking, particularly in the brown dwarf regime.
 
In a recent breakthrough, scaling laws derived from planetary dynamo calculations \citep{christensenAubert2006} were demonstrated to be empirically consistent with the magnetic field strengths measured for fully convective stars \citep{christensen2009}. This result argued for a single unifying principle that governs magnetic activity in rapidly rotating fully convective objects, spanning the mass range from stars to planets; specifically, that the energy flux available for generating the magnetic field sets the field strength. This principle states that the magnetic energy in these objects should scale approximately as $\propto \langle\rho\rangle^{1/3}q_0^{2/3}$, where $\langle\rho\rangle$ is the mean density in the dynamo region and  $q_0$ is the bolometric flux. However, while this scaling law appears consistent with magnetic field measurements for Solar System planets and fully convective stars, data from the orders of magnitude mass gap occupied by rapidly rotating brown dwarfs and massive extrasolar planets are required to validate this principle. 
 
Traditionally, the Zeeman effect has been one of the most powerful means to measure the strength, filling factor, and even large-scale field topology of stellar magnetic fields, including those of fully convective stars. Zeeman broadening of atomic lines such as Fe~I has been successfully used to recover the large-scale field topologies of active mid-M dwarfs, confirming that the high levels of coronal and chromospheric activity observed for these stars is indeed associated with strong magnetic fields (typically a few kG) covering a large fraction of the photosphere (with filling factors as high as $\sim$50\%) \citep{johnskrullValenti1996}.  Zeeman Doppler imaging, involving time-resolved high-resolution spectropolarimetry, has been successfully applied to mid- and late-M dwarf stars, both above and below the fully convective boundary \citep{donati2006}. In some cases, strong large-scale poloidal fields are identified while in other cases weak large-scale fields with strong higher order components are found \citep{morin2010}, suggesting that a bistable dynamo may operate in the very low-mass regime. Probing to even cooler temperatures,  \cite{reinersBasri2007} were able to use Zeeman broadening of magnetically sensitive molecular lines, such as FeH, to constrain the average surface magnetic fluxes of objects as late as M9. While these methods have been successful, a robust detection of Zeeman broadening has not been established for objects cooler than late M, as rapid rotational broadening causes blending of the desired molecular lines \citep{reinersBasri2006}.
    
In the last decade, observations of the radio emission from low mass stars and brown dwarfs have opened a new window on magnetic activity in this regime. While the initial detection of quiescent emission from $\sim$10\% of targets \citep{berger2006}, possibly consistent with incoherent gyrosynchrotron emission, was itself anomalous \citep{berger2001}, the later confirmation of a second component to the radio emission, manifested as periodic pulsar-like bursts of $100\%$ circularly polarized emission, was even more unexpected \citep{hallinan2006, hallinan2007}. This emission is attributed to the electron cyclotron maser instability, and is of the same nature as the auroral emission produced by the magnetic planets in our Solar System via magnetosphere-ionosphere coupling. However, unlike the planets, where auroral radio emission is powered by interactions with the solar wind, orbiting satellites, and co-rotation breakdown, the nature of the electrodynamic engine powering auroral activity in ultracool dwarfs remains unclear \citep{hallinan2015}. 
    
What is clear is that electron cyclotron maser (ECM) emission is a very powerful tool for measuring magnetic fields. Produced at the electron cyclotron fundamental frequency $\nu_{\mathrm{MHz}} \sim 2.8 \times B_{\mathrm{Gauss}}$ \citep[and references therein]{treumann2006}, it allows for very accurate measurements of local magnetic field strengths and rotation periods, and it has provided some of the first confirmations of kilogauss fields for late M and L dwarfs \citep{burgasser2005_radio, hallinan2006, hallinan2007, hallinan2008, berger2009}.  Indeed, radio observations have been the only method thus far capable of direct magnetic field measurements for L dwarfs.  Examining magnetic dynamo action in the mass gap between planets and stars requires radio data. 
    
Over a dozen low mass stars and brown dwarfs, ranging in spectral type from M7-L5, have been found to be radio sources in the last decade \citep{berger2001, berger2002, burgasser2005_radio, berger2006, phanbao2007, antonova2007, mclean2012, burgasser2013, williams2014, burgasser2015b}. A subset of these objects have been the subject of lengthy follow-up campaigns that have revealed the presence of 100\% circularly polarized, periodic pulses, with the pulse period typically 2--3 hours and consistent with rotation \citep{hallinan2006, hallinan2007, hallinan2008, berger2009}. More recently, magnetic field measurements have been extended much further, with the detection of the coolest radio brown dwarf yet detected, the T6.5 dwarf 2MASS J10475385+2124234 (hereafter 2M1047) by \cite{route2012}. They observed individual radio pulses from this object in multiple short duration observations at 4.75~GHz with the Arecibo observatory, resulting in a confirmed magnetic field strength of at least 1.7~kG near the surface of this extremely cool ($\sim$900~K) object. The results of \citet{route2012} highlight the unique capability of radio observations to measure magnetic fields in the critical L and T dwarf regime and demonstrates that the latest-type brown dwarfs can in fact host $\sim$kG field strengths.
    
However, this single detection came at substantial expense. In previous surveys totaling $\sim$60 L6 or later type objects, only one was detected \citep{antonova2013, route2013}, demonstrating that  previous selection strategies (largely volume-limited) have been inefficient. Motivated by the radio detection of 2M1047, we present a pilot survey of 6 objects ranging in spectral type L7.5--T6.5, including the previously detected T6.5 dwarf 2M1047.  We selected our targets using a new strategy, described in \S \ref{sec.selection}.  We measure magnetic field strengths of the coolest brown dwarfs using auroral radio emission, and we study implications on fully convective magnetic dynamo theory.

\section{Target Selection Strategy}\label{sec.selection}

In a departure from previous surveys, we have selected our objects for tracers of auroral emission at other wavelengths. This selection strategy is motivated by recent work by \citet{hallinan2015} linking periodic auroral radio emission to H$\alpha$ emission and optical broadband variability, as well as corroborating evidence that most radio-pulsing ultracool dwarfs exhibit weak H$\alpha$ emission and/or optical/IR variability. 

H$\alpha$ and X-ray emission have been known for decades to scale as power laws of increasing surface rotation or decreasing Rossby number ($Ro \sim P/\tau_c$, where $P$ is the stellar rotation period and $\tau_c$ is the convective turnover time) for main sequence F through mid-M stars, until around $Ro\sim 0.1$, when the activity-rotation scaling appears to saturate at a constant $\log L_{\mathrm{X, H\alpha}} / L_{\mathrm{bol}}$ \citep{pallavicini1981, soderblom1993, stauffer1994, delfosse1998, pizzolato2003, reinersBasriBrowning2009, mclean2012}. Additionally, flaring and quiescent radio emission observed in dwarf stars have been attributed to magnetic activity in the corona \citep{white1989, drake1989}, and in fact, X-ray and radio luminosities for magnetically active stars are tightly correlated on the G{\"u}del-Benz relation, spanning 5--6 orders of magnitude and including F through M stars and solar flares \citep{gudelBenz1993}.  The G{\"u}del-Benz relation holds for active stars independent of age, spectral class, binarity, or rotation period.  It suggests that coronal heating and particle acceleration via magnetic fields are related processes \citep[and references therein]{forbrich2011}. 

However, beyond $\gtrsim$M7, magnetic activity trends appear to diverge. L and T dwarfs regardless of age appear to be fast rotators \citep{reinersBasri2008}, suggesting that they do not spin down with age like M dwarfs.  $\gtrsim$M7 dwarfs also exhibit systematically weaker H$\alpha$ emission despite being fast rotators, while $L_{\mathrm{X}} / L_{\mathrm{bol}}$ decreases with increasing $v\sin i$ or decreasing $Ro$ \citep{mohantyBasri2003, reinersBasri2008, reinersBasri2010, berger2010, mclean2012}.  In a similar vein, the G{\"u}del-Benz relation appears to break down for objects later than M7 due to a suppression of X-ray luminosities rather than radio luminosities, even when taking activity-rotation saturation into account \citep{berger2010, williams2014}, suggesting that magnetic activity in L and T dwarfs is no longer dominated by rotation \citep{cook2014, williams2014}.  Although radio, H$\alpha$, and X-ray luminosities do not necessarily scale with magnetic field strength, their continued emission requires magnetic fields even at very low masses.  Zeeman broadening and Zeeman Doppler imaging studies referenced in \S \ref{sec.intro} confirm that $\sim$kG fields persist in dwarfs as late as M7. In light of such magnetics fields, a simple explanation for the observed activity breakdowns may be the decoupling of magnetic fields from increasingly neutral atmospheres \citep{mohanty2002}.  
  
However, clearly nonthermal heating of the upper atmospheres of ultracool dwarfs is commonplace and sustained.  The breakdown of activity trends in late-type dwarfs indicates that the persistence of H$\alpha$, X-ray, and radio emission perhaps reflects a departure from the standard chromospheric heating picture where magnetic fields locally interact with hotter and less neutral atmospheres. Instead, activity may be externally powered via auroral current systems such as magnetosphere-ionosphere (M-I) coupling currents, giving rise to auroral activity \citep{schrijver2011, nichols2012, hallinan2015}.  M-I coupling has been confirmed as a source of power for Jovian, Saturnian, and terrestrial auroral emissions \citep[and references therein]{hill1979, nichols2012, bagenal2014, badman2015}.

Recently, \citet{hallinan2015} have established that radio emission may only be one manifestation of auroral activity, as is observed for the planets in our Solar System. These authors have shown that the M8.5 dwarf LSR J1835+3259 is simultaneously variable with the same periodicity in broadband optical, Balmer line, and pulsed radio emission.  The radio and H$\alpha$ luminosities, together with the synchronized variability, are consistent with the emission in all bands being powered by the same auroral currents. \citet{hallinan2015} also postulated that there may be a causal relationship between auroral currents and some examples of the infrared variability (weather) observed for L and T dwarfs, though they presented no empirical data to support this hypothesis. 

Such synchronized multiwavelength emission has been previously observed in other radio brown dwarfs. TVLM~513-46546 (M8.5) exhibited anticorrelated Sloan-$g'$ and Sloan-$i'$ lightcurves, which \citet{littlefair2008} attributed to cloud phenomena, and H$\alpha$ emission from 2MASSW~J0746425+200032 (L0+L1.5) was variable with the same periodicity as its pulsed radio emission but at a 1/4-phase lag \citep{berger2009}.  In fact, all but one of the known radio-pulsing ultracool dwarfs also exhibit H$\alpha$ emission and several are also confirmed optical/IR variables \citep[and references therein]{tinneyReid1998, delfosse2001, basri2001, hall2002, reid2002, mohantyBasri2003, fuhrmeisterSchmitt2004, lane2007, schmidt2007, littlefair2008, berger2009, berger2010, reinersBasri2010, harding2013, antonova2013, burgasser2015a}.  

Motivated by the above discussion, we strongly bias our samples for auroral activity by targeting only those dwarfs in this spectral range known to exhibit H$\alpha$ emission and/or optical/IR variability.

\section{Targets}\label{sec.Targets}

\setlength{\tabcolsep}{0.05in}
\begin{deluxetable*}{lllllllll}[htp]
\tablecaption{Survey Targets \label{table:properties}}
\tablehead{
	\colhead{Object Name}               &
	\colhead{Abbrev.}                   & 
	\colhead{SpT}                       &
	\colhead{Parallax}                  &
	\colhead{Distance}                  &
	\colhead{$\mu_{\alpha}\cos\delta$}  &
	\colhead{$\mu_{\delta}$}            &
	\colhead{Notes}                     &
	\colhead{Ref.}                  
	    \\
	\colhead{}                          & 	
	\colhead{Name}                      & 
	\colhead{}                          &
	\colhead{(mas)}                     &	
	\colhead{(pc)}                      &
	\colhead{(mas/yr)}                  &
	\colhead{(mas/yr)}                  &
	\colhead{}                          &
	\colhead{}                      
}
\startdata
2MASS~10475385+2124234 & 2M1047   & T6.5 	& 94.73$\pm$3.81 	& 10.56$\pm$0.52      & -1714\phd$\pm$7       & -489\phd$\pm$4			& H$\alpha$, detected prior									& 1-7  			\\
2MASS~01365662+0933473 & SIMP0136 & T2.5 	& \nodata		& \phn6.0\phn$\pm$0.4 & \phm{-}1241\phd$\pm$9 & \phn\phn-4\phd$\pm$10		& IR variability												& 8-10 			\\
2MASS~10430758+2225236 & 2M1043   & L8   	& \nodata   	 	& 16.4\phn$\pm$3.2    & -134.7$\pm$11.6       & -5.7\phn$\pm$17.0  		& H$\alpha$ emission											& 11-13 			\\
2MASS~12373919+6526148 & 2M1237   & T6.5 	& 96.07$\pm$4.78	& 10.42$\pm$0.52      & -1002\phd$\pm$8       & -525\phd$\pm$6 			& H$\alpha$, IR var?\tablenotemark{a}							& 1 3 4 14-16	\\
SDSS~J12545393-0122474 & SDSS1254 & T2		& 75.71$\pm$2.88 	& 13.21$\pm$0.50      & \phn-479\phd$\pm$3    & \phm{-}130\phd$\pm$2  	& H$\alpha$, IR var?\tablenotemark{b}, binary?\tablenotemark{c}	& 17 3 4 18-26	\\
SDSS~04234858-0414035  & SDSS0423 & L7\tablenotemark{d} &  65.93$\pm$1.7 	& 15.17$\pm$0.39      & \phn-331\phd$\pm$49   & \phm{-}\phn76\phd$\pm$11	& H$\alpha$, IR var, binary\tablenotemark{d} 					& 19 27-33
\enddata
\tablenotetext{a}{(14) found no evidence of J-band variability whereas (16) reported variability at a level below the detection limits of (14) }
\tablenotetext{b}{(22), (23), (24) found no IR variability in SDSS1254 above the $\sim5-20$~mmag level, whereas (20) and (21) reported `significant' J-band and spectroscopic variability, respectively. }
\tablenotetext{c}{See (25), (26) and \S \ref{sec.physParam} and \S \ref{sec.Targets}  for further discussion about possible multiplicity in SDSS1254.  }
\tablenotetext{d}{SDSS0423 has a known binary companion of spectral type T2.5 and orbital separation 0$\farcs$16 (31, 32, 33). }
\tablerefs{ 
(1) \cite{burgasser1999};
(2) \cite{burgasser2006a};
(3) \cite{vrba2004};
(4) \cite{burgasser2003_redOpticalData}; 
(5) \cite{route2012};
(6) \cite{williams2013};
(7) \cite{williams2015};  
(8) \cite{artigau2006}; 
(9) \cite{artigau2009};
(10) \cite{apai2013};
(11) \cite{cruz2007};
(12) \cite{schmidt2010};
(13) \cite{pinedaInPrep};   
(14) \cite{burgasser2002b};
(15) \cite{burgasser2000b};
(16) \cite{artigau2003};  
(17) \cite{leggett2000};
(18) \cite{burgasser2002a};
(19) \cite{geballe2002};
(20) \cite{artigau2003};
(21) \cite{goldman2008}
(22) \cite{koen2004};
(23) \cite{girardin2013};
(24) \cite{radigan2014a};
(25) \cite{burgasser2007};
(26) \cite{cushing2008}; 
(27) \cite{cruz2003};
(28) \cite{kirkpatrick2008};
(29) \cite{enoch2003};
(30) \cite{clarke2008};
(31) \cite{carson2011};
(32) \cite{burgasser2005b};
(33) \cite{burgasser2006b}}
\end{deluxetable*}

\textbf{2MASS~10475385+2124234}.
Discovered by \cite{burgasser1999}, 2M1047 was later classified as a T6.5 brown dwarf by \cite{burgasser2006a}. \cite{burgasser2003_redOpticalData} detected weak H$\alpha$ emission at the 2.2$\sigma$ level with a flux of $f_{\mathrm{H}\alpha}= 5.9 \pm 2.7 \times 10^{-18}$~ergs~cm$^{-2}$~s$^{-1}$. In 2012, 2M1047 became the first T-dwarf detected in the radio, when \cite{route2012} detected highly circularly polarized ($\gtrsim$72\%) and bright flares at 4.75~GHz with $\sim$1.3--2.7~mJy peak flux densities using the Arecibo telescope.  Until this study, it has remained the only radio-detected $\geq$L6 dwarf.  A follow-up study by \cite{williams2013} at 5.8~GHz using the VLA found quasi-quiescent radio emission from this source with a flux density of $16.5 \pm 5.1~\mu$Jy.  \cite{williams2015} confirmed quiescent emission for 2M1047, measuring a flux density of $9.3 \pm 1.5~\mu$Jy and  $1.1 \pm 1.5~\mu$Jy at 6--10~GHz for Stokes I and V, respectively, with low circular polarization ($\lesssim$28\%).  They also detected highly left-circularly polarized pulses ($\sim$50--100\%) with a periodicity of $\sim$1.77 hours up through 10~GHz.  We include 2M1047 in our survey as a known quiescently emitting source and to examine long-term variability.

\textbf{SIMP~J01365662+0933473}.
SIMP0136 was discovered and classified as a T2.5 dwarf by \cite{artigau2006}.  In a follow-up study, \cite{artigau2009} reported J- and K$_s$-band photometric variability, with a peak-to-peak amplitude $\Delta J \sim 50$~mmag, an amplitude ratio of $\Delta K_s / \Delta J=0.48 \pm 0.06$, and a period $P=2.3895\pm0.0005$~hr.  This was the first clearly periodic and high-amplitude detection of IR variability in a T-dwarf.  They attributed the variability to clouds that are $\sim$100~K colder than a surrounding cloud-free atmosphere in the brown dwarf.  Using HST spectral mapping, \cite{apai2013} found that models of low-temperature and thick clouds mixed with warmer and thin clouds can reproduce time-variable changes in the near-IR colors and spectra of SIMP0136, and they confirmed it had a stable variation period. 

\textbf{2MASS~10430758+2225236}.
2M1043 was discovered and classified as an unusually red L8 dwarf by \cite{cruz2007}, which they speculated could be attributed to an unresolved binary.  A follow-up study by \cite{reid2008} using the NICMOS N1C1 camera on the Hubble Space Telescope found that no binary companion to 2M1043 was resolved, for mass ratios $q > 0.2$ and angular separations $\theta>0\farcs3$.  In the discovery paper, the authors also tentatively report possible H$\alpha$ emission.

\textbf{2MASS~12373919+6526148}.
2M1237 was discovered by \cite{burgasser1999} using data from the Two Micron All-Sky Survey \citep{2mass2006} and classified as a T6.5 dwarf by \cite{burgasser2002b}.  \cite{burgasser2000b, burgasser2002b} reported abnormally bright and persistent yet variable H$\alpha$ emission, which was confirmed again by \cite{burgasser2003_redOpticalData}.  With fluxes ranging from  $f_{\mathrm{H}\alpha}\sim1$--10$\times 10^{-17}$~erg~cm$^{-2}$~s$^{-1}$, the H$\alpha$ luminosity is an order of magnitude higher than for any other T dwarf. \cite{burgasser2002b} found no evidence of short-term J-band variability and ruled out flaring as a possible variability mechanism.  In contrast, \cite{artigau2003} reported variability at $\Delta J \sim30$~mmag.  \cite{liebert2007} ruled out a massive companion or youthful chromospheric activity as additional possible H$\alpha$ variability mechanisms. 

\textbf{SDSS~J12545393-0122474}.
SDSS1254 was discovered by \cite{leggett2000} and independently classified as a T2 dwarf by both \cite{burgasser2002a} and \cite{geballe2002} and is the T2 spectral standard \citep{burgasser2006a}.  \cite{burgasser2003_redOpticalData} reported weak H$\alpha$ emission with flux $f_{\mathrm{H}\alpha}= 7.5 \pm 2.5\times 10^{-18}$~erg cm$^{-2}$ s$^{-1}$. \cite{artigau2003} reported 45$\pm$2~mmag J-band and 23$\pm$4~mmag H-band variability, and similarly, \cite{goldman2008} report variable spectral features at $0.997-1.13~\mu$m, with upper limits in the peak-to-peak flux variability calculated at the $\sim$4--60\% levels. In contrast, \cite{koen2004} found no evidence of variability above the 7, 6, and 10~mmag levels for J, H, and K$_\mathrm{s}$ bands during a $\sim$4-hour observation, and \cite{girardin2013} found no evidence of J-band variability above 5~mmag.  We note here that SDSS1254 appears to be sufficiently overluminous for its spectral type that it may in fact be an as-yet unresolved tight binary system \citep{burgasser2007, cushing2008}.

\textbf{SDSS~04234858-0414035}.
SDSS0423 was identified by \cite{geballe2002} using data from the Sloan Digital Sky Survey \citep{sdss}. The authors classified it as a T0 dwarf on the basis of its infrared spectrum.  However, using its optical spectrum, \cite{cruz2003} classified it as an L7.5.  \cite{burgasser2005b} showed that it is in fact a binary system of two brown dwarfs with spectral types L6$\pm$1 and T2$\pm$1, consistent with the previous classifications.  Both \citet{burgasser2005b} and \citet{carson2011} reported the angular separation of the binary to be 0$\farcs$16, which we cannot resolve with our observations.  For the purposes of comparing our magnetic field measurements to previous models, we adopt a conservative L7.5 classification. Monitoring in K$_{\mathrm{s}}$ band by \cite{enoch2003} yielded only a possible detection of variability, whereas \cite{clarke2008} reported J-band photometric variability with a peak-to-peak amplitude of $8.0\pm0.8$~mmag with a period of $2\pm0.4$~hr.  SDSS0423 is additionally one of only a handful of late L/T-dwarfs to exhibit H$\alpha$ emission, for which \cite{kirkpatrick2008} reported an equivalent width of 3~\AA.

\section{Observations}\label{sec.Observations}

\setlength{\tabcolsep}{0.05in}
\begin{deluxetable*}{lccclccclc}[htp]
\tablecaption{Summary of observations\label{table:obs}}
\tablehead{
	\colhead{}                  &
	\colhead{}                  &
	\colhead{Obs.}              &
	\colhead{Obs.}              &
	\colhead{Time on}           &
	\colhead{VLA}               &
	\colhead{Synthesized Beam}  &
	\colhead{}                  &
	\colhead{Phase}             &
	\colhead{Flux}          
	    \\
	\colhead{Object}            &
	\colhead{Band}              &
	\colhead{Date}              &
	\colhead{Block}             &
	\colhead{Source}            & 
	\colhead{Configuration}     &
	\colhead{Dimensions}        &
	\colhead{RMS}               &
	\colhead{Calibrator}        &
	\colhead{Calibrator}    
	    \\
	\colhead{}                  &
	\colhead{(GHz)}             &
	\colhead{(2013)}            &
	\colhead{(h)}               & 
	\colhead{(s)}               & 
	\colhead{}                  &
	\colhead{(arcsec $\times$ arcsec)}            &
	\colhead{($\mathrm{\mu}$Jy)}& 
	\colhead{}                  &
	\colhead{}                  
}
\startdata
2M1047                      & 4.0--8.0 & 05/19 & 4.0 & 12745   & DnC & \phn9.21 $\times$ 3.02  & 3.1     & J1051+2119 & 3C286 \\[8pt]
SIMP0136                    & 4.0--8.0 & 05/18 & 4.0 & 12995   & DnC & \phn8.64 $\times$ 3.10  & 5.4     & J0203+1134 & 3C147 \\[8pt]

2M1043                      & 4.0--8.0 & 05/25 & 4.0 & 13042.5 & DnC & \phn10.0 $\times$ 5.5   & 2.0     & J1051+2119 & 3C286 \\
                            & 4.0--8.0 & 05/27 & 2.0 & 5825    & DnC & \phn9.82 $\times$ 5.47  & 4.9     & J1051+2119 & 3C286 \\[8pt] 
         
2M1237                      & 4.0--8.0 & 05/21 & 2.0 & 5712.5  & DnC & \phn8.22 $\times$ 3.70  & 2.8     & J1313+6735 & 3C286 \\[8pt]

SDSS1254\,\tablenotemark{a} & 4.0--8.0 & 05-19 & 2.0 & 5685    & DnC & \phn9.70 $\times$ 3.55  & 4.0     & J1246-0730 & 3C286 \\
                            & 4.0--8.0 & 05/26 & 2.0 &         & DnC &         \nodata         & \nodata & J1246-0730 & 3C286 \\[8pt]
         
SDSS0423                    & 4.0--8.0 & 08/30 & 4.0 & 13102.5 & C   & \phn4.91 $\times$ 3.37  & 3.2     & J0423-0120 & 3C147 \\
                            & 4.0--8.0 & 05/26 & 2.0 & 5907.5  & DnC & 11.52 $\times$ 5.96     & 4.0     & J0423-0120 & 3C147 \\   
                            & 4.0--8.0 & 05/25 & 2.0 & 5925    & DnC & 12.92 $\times$ 9.11     & 3.5     & J0423-0120 & 3C147
\enddata
\tablenotetext{a}{Unable to successfully calibrate measurement set taken on 2013-05-26 due to excessive noise.}
\end{deluxetable*}

We observed 6 objects spanning spectral range L7.5-T6.5 with the full VLA array in C-band (4--8 GHz), using the WIDAR correlator in 3-bit observing mode for 4~GHz bandwidth observations, in time blocks of 2 or 4 hours for 28 total program hours.  Observations were performed between March and August 2013, during DnC and C configurations. We summarize target properties and observations in Table \ref{table:properties} and Table \ref{table:obs}, respectively.

\subsection{Calibrations\label{section:calibrations}}
We calibrated our measurement sets using standard VLA flux calibrators 3C286 and 3C147 and nearby phase calibrators.  After initially processing raw measurement sets with the VLA Calibration Pipeline, we manually flagged remaining RFI.  Typical full-bandwidth sensitivity at DnC configuration for 2 hours on source in C-band is 3\,$\mu$Jy. Typical 3-bit observations reach an absolute flux calibration accuracy of $\sim$5\%. We obtained absolute flux by bootstrapping flux densities with standard VLA flux calibrators.  Flux calibration accuracy may be reduced and result in systematically offset flux densities when gain calibrations interpolated from the phase calibrator are not sufficient to correct for the variation of gain phases with time. To account for this, our observations alternated between a nearby phase calibrator and the target source with  typical cycle times of 30 minutes, and we obtained gain solutions for the phase calibrators that varied slowly and smoothly over time, suggesting that this source of error is negligible. 

\pagebreak
\subsection{Source Motion\label{section:motion}}
The expected positions of the sources were determined using 2MASS coordinates \citep{2mass2006} and corrected for proper motion, provided in Table \ref{table:properties}.  Sources had moved by as much as 0$\farcs$8 due to proper motion during our observing program, in comparison to synthesized beam resolutions of at least a few arcseconds.  Orbital motion corrections were not necessary for SDSS0423, a known binary with an orbital separation 0$\farcs$16.  We compared the expected coordinates of our objects to their measured position and found that all objects were well within a synthensized beam of their predicted locations.

\section{Results}\label{sec.Results}
\subsection{Image Detections}
\begin{figure*}
\epsscale{1.1}
\plotone{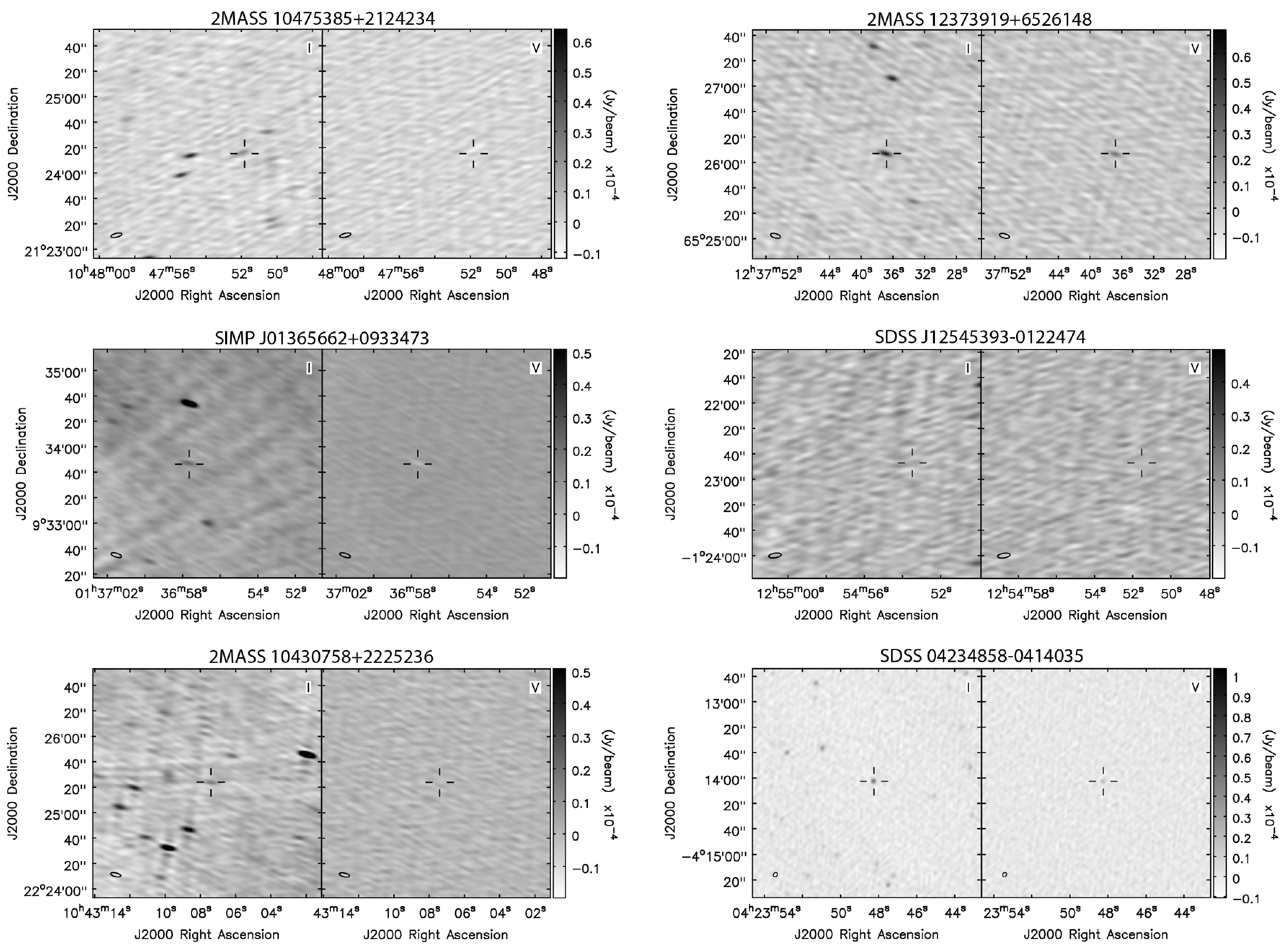}
\caption{\label{fig:imagesI} Stokes I (left) and V (right) images of all objects.  Ellipse depicts synthesized beam.  Measurement sets for objects with multiple observing blocks were concatenated prior to imaging.  Sources were detected at the proper motion-corrected location for all objects except for SDSS1254.}
\end{figure*}

We combined measurement sets for objects with multiple observing blocks using the CASA \texttt{concat} routine and then produced Stokes I and Stokes V images of each object (total and circularly polarized intensities, respectively) with the CASA \texttt{clean} routine, modeling the sky emission frequency dependence with 2 terms and using Brigg's weighting with the robustness parameter set to 0.0, which we found resulted in a good trade-off between resolution and sensitivity for our observations. We searched for a point source at the proper motion-corrected coordinates of each target.  Figure \ref{fig:imagesI} shows the Stokes I and Stokes V images for all objects. 

In contrast to previous surveys, all but one of our six targets were detected in Stokes I, with signal-to-noise ratios (SNR) ranging from 4.9 to 24.6 in the mean Stokes I flux density.  Table \ref{table:timeseries} gives the measured mean flux density and rms noise of each detected (SNR $\ge$ 3) source.  Flux densities and source positions were determined by fitting an elliptical Gaussian point source to the cleaned image of each object at its predicted coordinates, using the CASA task \texttt{imfit}.  For the one undetected target, SDSS1254, we provide the measured mean Stokes I flux density and rms noise at the expected position of the source.

\newpage
\subsection{Timeseries Pulse Detections \& \\ Magnetic Field Strengths} \label{subsec.Timeseries}
\begin{figure*}
\epsscale{1.1}
\plotone{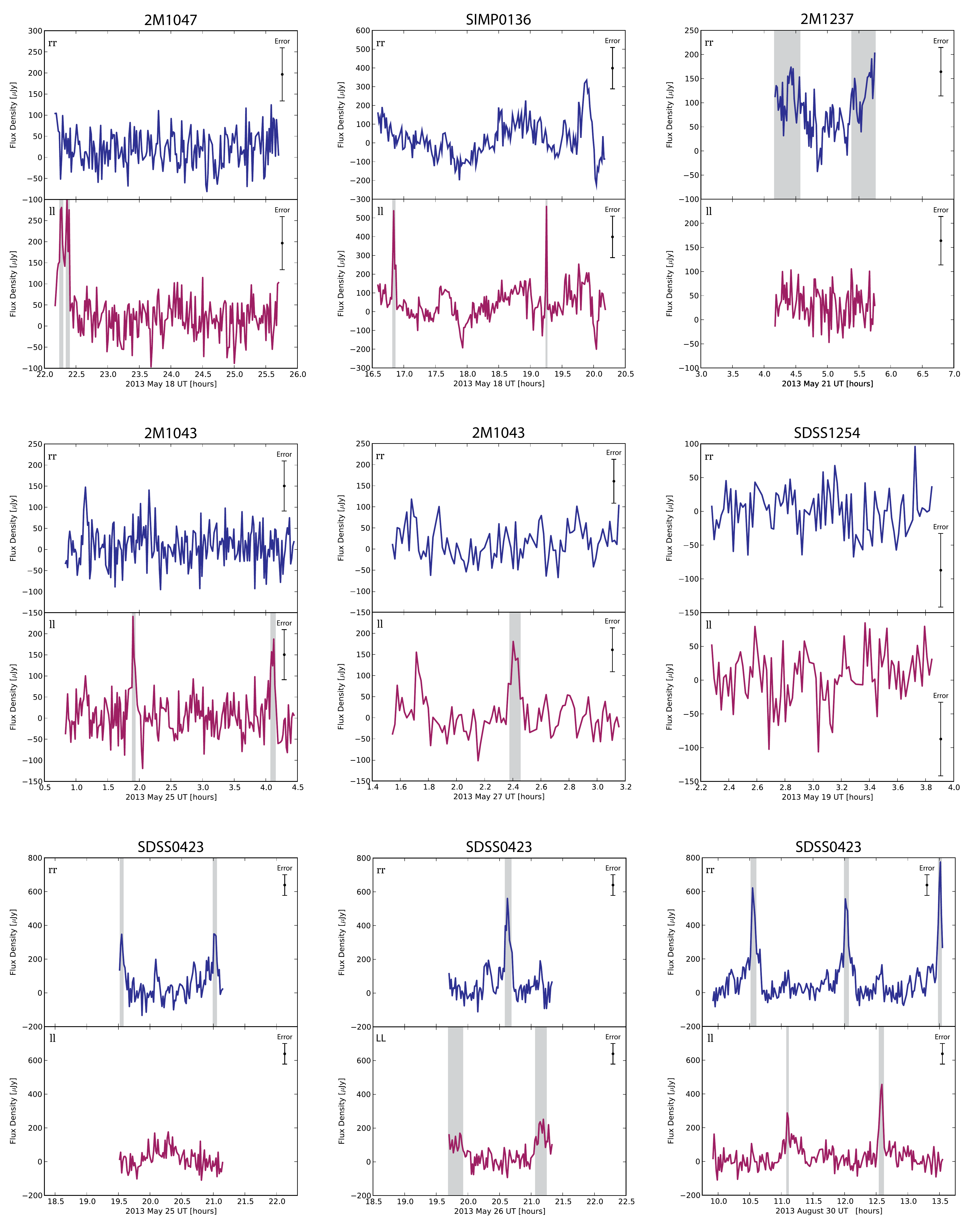}
\caption{\label{fig:alltimeseries} Timeseries of rr- and ll-correlated (blue for right circularly polarized and red for left circularly polarized, respectively) flux densities for all calibrated measurement sets.  Axes scales are constant for timeseries for objects with multiple observing blocks.  For presentational clarity, data is averaged over 60s intervals; time interval for raw data was 5s seconds and all analysis was done with data averaged over 10s.  Black error bars represent rms noise obtained in images and scaled to time bin lengths for a single correlation.  Grey regions indicate FWHM of pulses with peak flux density $\geq$3.0, and all pulses have been verified with imaging. Total intensity is given by the Stokes I flux density, where I~=~(rr+ll)/2.   Circularly polarized intensity is given by the Stokes V flux density, where V~=~(rr-ll)/2.  }
\end{figure*}

\begin{figure*}
\epsscale{1}
\plotone{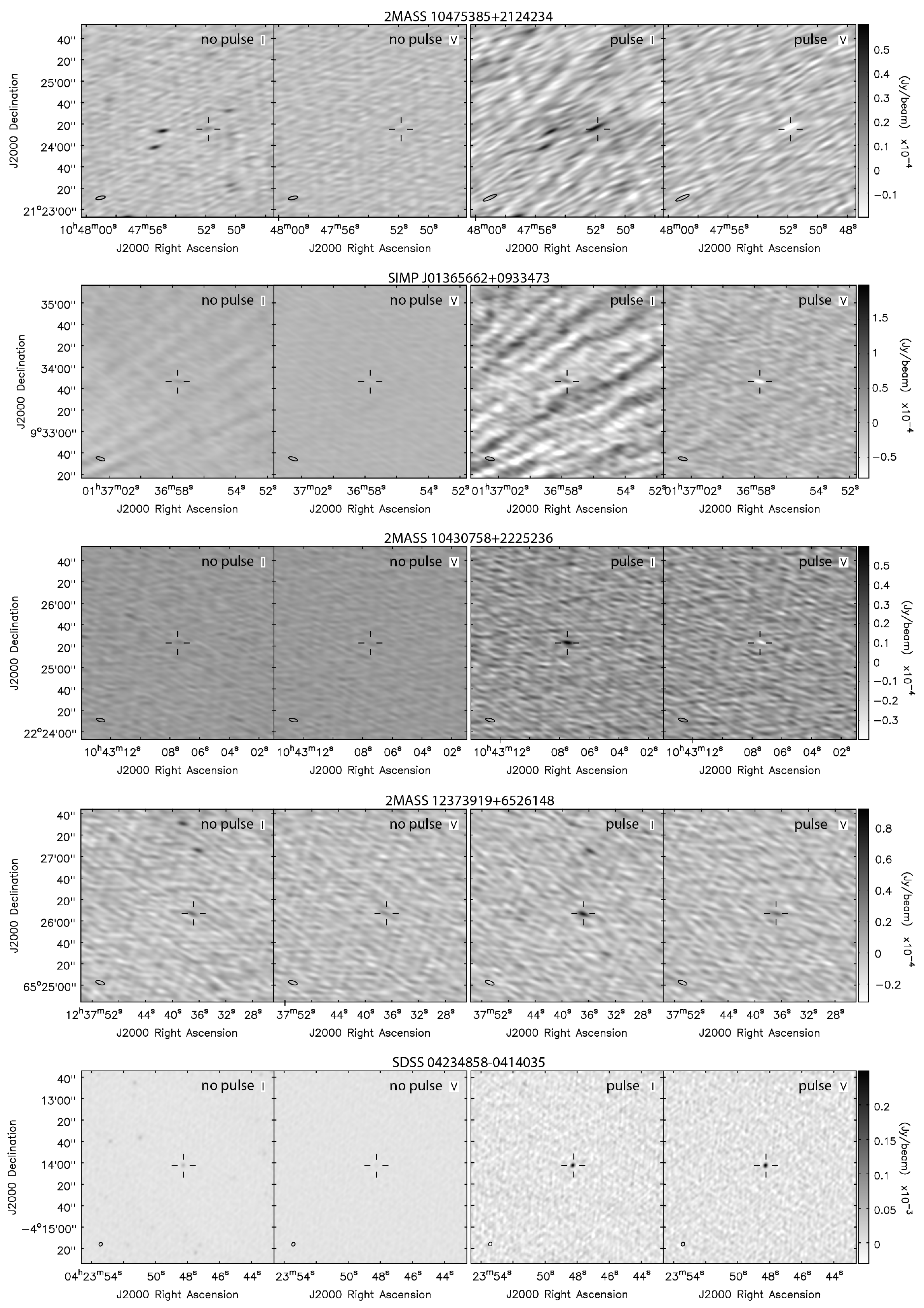}
\caption{Stokes I and Stokes V flux densities for pulsed and quiescent emission.  Pulsed emission for 2M1237 is averaged only over the later pulse, and SDSS0423 pulsed emission is averaged over the rr pulses only.   \label{fig:quiescentEmission} }
\end{figure*}

\setlength{\tabcolsep}{0.05in}
\begin{deluxetable*}{lcccccccccc}[htp]
\tablecaption{Imaging and Timeseries Results\label{table:timeseries}}
\tablehead{
	\colhead{}								&
	\colhead{Position}						&
	\colhead{Mean }							&
	\colhead{Pulse}							&
	\colhead{Pulse}							&
	\colhead{Pulse}							&
	\colhead{}								&
	\colhead{Pulse}							&
	\colhead{Quiescent}						&
	\colhead{}								&
	\colhead{Quiescent}
	    \\
	\colhead{Object}						&
	\colhead{Offset\,\tablenotemark{a}}		&
	\colhead{Stokes I}						&
	\colhead{\#}							&
	\colhead{Stokes I}						&
	\colhead{Stokes V}						&
	\colhead{SNR}							&
	\colhead{Circ. Poln}					&
	\colhead{Stokes I}						&
	\colhead{SNR}							&	
	\colhead{Circ. Poln}
	    \\
	\colhead{}								&
	\colhead{(sigma)}						&
	\colhead{($\mu$Jy)}						&
	\colhead{}								&
	\colhead{($\mu$Jy)}						&
	\colhead{($\mu$Jy)}						&
	\colhead{(I,V)}							&
	\colhead{(\%)\,\tablenotemark{b}}		&
    \colhead{($\mu$Jy)}						&
	\colhead{}								&
	\colhead{(\%)\,\tablenotemark{b}}    
	}                              
\startdata
2M1047						& 1.46  & 26.8$\pm$3.1    & \phn1\phm{?}\,\phm{$^{c}$}& 123.0$\pm$21.0    	& \phn-95.0$\pm$15.0          & \phn5.9, \phn6.3 & -75.1$_{-14.9}^{+14.1}$     & 17.5$\pm$3.6 & 4.9 	& -40.6$_{-13.2}^{+23.4}$			\\[8pt]
SIMP0136						& 0.36  & 34.4$\pm$5.4    & \phn2\phm{?}\,\phm{$^{c}$}& $>$156.0$\pm$39.7\,\tablenotemark{c}    	& -233.0$\pm$24.9             & \phn3.9, \phn9.4 & -63.6\,\tablenotemark{d}    & 33.3$\pm$5.9 & 5.6 	& -1.2\,\tablenotemark{d}			\\[8pt]
2M1043						& 0.79  & 11.7$\pm$2.4    & \phn3\phm{?}\,\phm{$^{c}$}& \phn87.0$\pm$11.8 	& \phn-69.0$\pm$11.7          & \phn7.4, \phn5.9 & -77.9$_{-13.0}^{+15.1}$     & 16.3$\pm$2.5 & 6.5 	& -13.8$_{-15.9}^{+13.8}$			\\[8pt]
2M1237\,\tablenotemark{e}		& 2.91  & 64.7$\pm$3.7    & \phn2?\,\tablenotemark{f} & \phn83.3$\pm$7.6  	& \phn\phm{-}23.7$\pm$6.4\phn & \phn9.5, \phn3.7 & \phm{-}28.2$_{-7.5}^{+9.0}$ & 43.3$\pm$7.3 & 5.9 	& \phm{-}53.7$_{-14.6}^{+21.6}$	\\[8pt]
							&\nodata& \nodata         & \nodata                   & \phn81.7$\pm$8.8  	& \phm{-}\phn40.3$\pm$8.0\phn & \phn9.3, \phn5.0 & \phm{-}48.8$_{-9.7}^{+13.1}$& \nodata      & \nodata	& \nodata 						\\[8pt]
SDSS1254 					&\nodata& \phn3.3$\pm$4.0 & \phn0\phm{?}\,\phm{$^{c}$}& \nodata           	& \nodata                     & \nodata          & \nodata                     & \nodata      & \nodata	& \nodata						\\[8pt] 
SDSS0423\,\tablenotemark{g}	& 0.42  & 54.1$\pm$2.2    & 10\phm{?}\,\phm{$^{c}$}   & 225.4$\pm$12.4    	& 220.0$\pm$12.2              & 18.2, 18.0       & \phm{-}97.3$_{-9.0}^{+0.8}$ & 26.7$\pm$3.1 & 8.6	& \phm{-}14.4$_{-10.2}^{+11.5}$	\\[8pt]
							&\nodata& \nodata         &\nodata                    & 135.0$\pm$9.8\phn 	& \phn-67.1$\pm$7.9\phn       & 13.8, \phn8.5    & -49.4$_{-7.8}^{+6.1}$       & \nodata      & \nodata	& \nodata 
\enddata
\tablenotetext{a}{The distance between the measured and expected coordinates, divided by the amplitude of the error ellipse in the offset direction, using concatenated images for objects with multiple observing blocks. 2MASS coordinate uncertainties and our own measurement uncertainties were included in error analysis.}
\tablenotetext{b}{Reported polarization fractions are highest-likelihood values, given the measured Stokes I and Stokes V flux densities.  Uncertainties reflect the upper and lower bounds of the 68.27\% confidence intervals. Negative values indicate left circular polarization, and positive values indicate right circular polarization. }
\tablenotetext{c}{Challenges with field source subtraction result in an underestimate of the true Stokes I flux density.  Because circular polarization cannot exceed 100\%, the Stokes V flux density gives a lower bound to the true Stokes I flux density (see c).}
\tablenotetext{d}{We quote the lower bound of the 99.73\% confidence interval for the percent circular polarization of SIMP0136 due to an underestimated Stokes I flux density.}
\tablenotetext{e}{Due to the broadness of the two observed peaks in the rr timeseries of 2M1237, we report measurements separately for each peak.  The top measurement is for the earlier peak and the bottom measurement is for the later peak.}
\tablenotetext{f}{See \S \ref{subsec.Timeseries} for discussion}
\tablenotetext{g}{We observe two sets of pulses, six in the rr timeseries (top) and four in the ll timeseries (bottom). }
\end{deluxetable*}

We checked all targets for highly circularly polarized pulses in flux density to confirm the presence of ECM emission. Previous studies have searched for pulsed emission in Stokes I and V, but we have chosen to search for pulses in the rr and ll correlations (right- and left-circularly polarized, respectively) because signal to noise is a factor of $\sqrt{2}$ higher in cases where the pulsed emission is 100\% circularly polarized.  

Using the CASA plotting routine \texttt{plotms} to export the real UV visibilities averaged across all baselines, channels, and spectral windows of the rr and ll correlations, we created rr and ll timeseries for all measurement sets with time resolutions of 10s, 60s, and 600s at frequency ranges of 4--6~GHz, 6--8~GHz, and 4--8~GHz to check for frequency-dependent ECM emission cutoff.  We do not check for pulses at frequency resolutions smaller than 2~GHz due to signal-to-noise concerns.  Figure \ref{fig:alltimeseries} shows the 4--8~GHz timeseries for each object. 

Analysis of the timeseries shows significant evidence of at least one pulse for 2M1047, SIMP0136, SDSS0423, and 2M1043.  Additionally, 2M1237 appears to exhibit very broad pulses or strongly variable emission.  We confirm pulses by imaging right circularly polarized and/or left circularly polarized emission over the full width half maximum (FWHM) of each pulse and measuring integrated flux densities using the CASA routine \texttt{imfit} at the expected locations of our targets. We find that flux densities for imaged pulses are consistent with pulses observed in the timeseries within 3$\sigma$. For all objects except for 2M1237, we smooth our data over 60s, 90s, and 180s to measure the FWHM. We find that the FWHM is consistent within $\sim$30s, except for the earlier ll pulse on 08/30/2013 for SDSS0423; when the smoothing is extended to 180s, the narrow peak smears out into the broader bump, and the returned FWHM is accordingly broader.  For the purposes of measuring a mean pulsed flux, we use the narrower FWHM.  Because of the broad nature of the peaks for 2M1237, we smooth over 180s, 270s, and 540s and find that the FWHM is consistent within $\sim$450s.

We measure the mean pulsed Stokes I and V flux densities by imaging over all of the pulses with peak flux density $\geq3.0$ for each object and calculate the highest likelihood percent circular polarization of the mean pulsed flux, where negative and positive percentages correspond to left and right circular polarization, respectively.  We report uncertainties that correspond to the upper and lower limits of the 68.27\% confidence interval.  We find that in all cases except for the first peak in 2M1237, the pulsed emission is highly circularly polarized (48.8--97.3\%), consistent with ECM emission \citep{treumann2006}. 

We additionally check for quiescent emission by removing the full width of each pulse from our data and imaging the remaining emission.  We define the full width of the pulse as beginning and ending at the time bins nearest the pulse maximum that have flux densities less than or equal to the rms noise. We find that pulse widths for each object are consistent within $\sim$60s ($\sim$500s for 2M1237) for all smoothing resolutions, and we select the widest returned width when removing each pulse.  All objects with pulsed emission also exhibit quiescent emission with relatively low polarization fractions, except for 2M1237.  In contrast, SDSS1254, for which no pulse is observed, does not exhibit any detectable quiescent emission above the rms noise. We report the characteristics of the pulsed and quiescent emission in Table \ref{table:timeseries}. 

Searching for the FWHM of 2M1047 reveals an apparent double peak, similar to what \citet{williams2015} observe.  Based on the periodicity observed by \citet{williams2015}, we classify this object as having a single pulse. However, in measuring the mean pulsed flux densities, we treat it as a double pulse and average over the FWHM of each pulse.

Two extremely bright sources near SIMP0136 resulted in poor Stokes I field source subtraction, and our reported Stokes I flux density is certainly an underestimate of the true flux density.  We attempted to self-calibrate this field, but were only able to achieve $\sim$10\% improvement. Beam squint causes the nearby bright sources to also appear in Stokes V but with much lower flux densities, and we therefore consider the Stokes V flux density of SIMP0136 to be more accurate.  Because the degree of circular polarization cannot be greater than 100\%, the Stokes V flux density in fact gives a lower bound on the actual Stokes I flux density.  Due to its extremely bright Stokes V flux density, we conclude that pulses from SIMP0136 are highly circularly polarized. 

We note that our observations only tentatively suggest that we observe ECM emission from 2M1237.  Despite the broad nature of the peaks in 2M1237, it is possible that the timeseries in fact exhibits two pulses rather than simply being variable, with the broadness arising from a geometric effect. We report the flux densities and circular polarization fractions for each of the peaks in the 2M1237 rr timeseries, and we find that in fact the circular polarization fraction appears to vary from peak to peak, from $\sim$30\% to $\sim$50\% on a 2-hour time scale.  Some of the variability may arise from the incomplete phase coverage, such that the earlier peak is averaged down more than the later peak.  Whereas the other radio-detected objects all exhibit marked differences in polarization fractions between pulsed and quiescent emission, the `quiescent' emission from 2M1237 exhibits $\sim$50\% circular polarization, which is similar to what we observe in at least one of the peaks.  This could be consistent with a geometry in which the ECM-emitting region of the magnetosphere is always visible, which would also explain the broadness of the peaks. Additional monitoring of 2M1237 for full phase coverage is necessary to determine the nature of these peaks.

Three possibilities may account for why we do not observe a pulse from SDSS1254: (1) SDSS1254 does not produce ECM emission, (2) SDSS1254 produces ECM emission with a cutoff frequency lower than 4.0 GHz, or (3) we did not observe it during a pulse and the auroral activity is variable. Table \ref{table:timeseries} summarizes timeseries data for all objects.  All detected pulses extend into the 6.0--8.0~GHz band, indicating that observations at higher frequencies are required to detect and measure an emission cutoff.  We conservatively use the center of the top band, 7.0~GHz, to calculate corresponding lower bound maximum surface field strengths of 2.5~kG.

\section{Discussion}\label{sec.Discussion}
\subsection{Auroral Radio Emission as a Precise Tool for Magnetic Field Measurement}
Auroral ECM emission from the planets in our Solar System is produced very close to the fundamental electron cyclotron frequency local to the source region. Though intrinsically narrow-band ($\Delta \nu \ll \nu$), the emission can be detected over a wide range of frequencies, as the process operates efficiently over a range of heights above the planetary surface, which maps to a range of field strengths. Taking the Jovian case as an example, auroral radio emission is detected from 10~kHz to 40~MHz frequencies, with the lowest frequency emission originating in source regions out to $>$5~R$_{\mathrm{Jup}}$, and the high frequency emission corresponding to the highest strength magnetic field regions (14~Gauss) just above the atmosphere in the auroral polar regions in the northern hemisphere \citep{zarka1998}. Observed remotely, independent of knowledge of the source region or the electrodynamic engine powering the auroral currents, the high frequency cut-off of this emission provides a good means to determine the maximum magnetic field strength in the magnetospheres of the magnetized planets.

We propose to utilize the highly circularly polarized component of the radio emission detected from our sample of cool brown dwarfs to similarly constrain the maximum magnetic field strengths in their magnetospheres, with a view to constraining the dynamo mechanism at work in their interiors.  In the absence of a clear cutoff in emission, we note that any detection can be equated to a robust lower limit on a maximum surface magnetic field strength. While the detection of such ECM emission provides exquisite measurement of local magnetic field strengths at the source of the radio emission, this must be translated to global parameters of particular use to dynamo modeling. Similarly, care must be taken in comparing these measurements with magnetic field measurements previously obtained for higher mass objects via Zeeman splitting/broadening and Zeeman Doppler imaging, as they are measuring distinct but complementary properties of the magnetic field. We address these issues in $\S$ \ref{sec.formalism}.

The ECM emission from our sample is detected across the entire band of our observations, which spans 4--8~GHz. Thus, in the absence of a clear cut-off in the emission, we can place a lower limit to the maximum surface magnetic field strength of 2.5~kG for all of our detected sample.  This assumes the emission is produced at the fundamental electron cyclotron frequency, rather than a higher harmonic, as is the case for Solar System planets. Electron cyclotron maser emission at higher harmonics has been invoked to explain coherent radio bursts from the Sun and active stars, where the coronal density is such that second-harmonic cyclotron absorption may prevent escape of emission at the fundamental frequency. Indeed, it has been shown that emission at the second and higher harmonic can dominate when the ratio of the plasma frequency to the electron cyclotron frequency exceeds $\sim$0.3 \citep{winglee1985}. However, in the case of our sample, this would require a local plasma density of $\sim$10$^{11}$~cm$^{-3}$, more indicative of hot stellar coronae than the cool neutral atmospheres of late L and T dwarfs, motivating the assumption of emission at the fundamental electron cyclotron frequency. 

To best inform our comparison of our results to dynamo models, we also estimate the relevant physical parameters for our brown dwarfs, as discussed in $\S$ \ref{sec.physParam}.

\subsection{Estimating Physical Parameters of Brown Dwarfs}\label{sec.physParam}
\setlength{\tabcolsep}{0.05in}
\begin{deluxetable*}{llllllllll}[htp]
\tablecaption{Brown Dwarf Physical Parameters\label{table:physparam}}
\tablehead{
	\colhead{}												&
	\colhead{}												&
	\colhead{}												&
	\colhead{}												&
	\colhead{}												&
	\colhead{} 												&
	\colhead{Adopted}										&
	\colhead{Adopted}										&
  	\colhead{Adopted}										&
	\colhead{Adopted}                                         
	\\  
	\colhead{Object}										&
	\colhead{SpT}											&
	\colhead{T$_{\mathrm{eff}}$\,\tablenotemark{a}}			&
	\colhead{log~g\,\tablenotemark{a}}						&
	\colhead{Age\,\tablenotemark{b}}						&
	\colhead{Mass\,\tablenotemark{b}} 						&
	\colhead{T$_{\mathrm{eff}}$\,\tablenotemark{c}}			&
	\colhead{log~g\,\tablenotemark{c}}						&
  	\colhead{Age\,\tablenotemark{c}}						&
	\colhead{Mass\,\tablenotemark{c}}                                         
	\\  
	\colhead{}												&
	\colhead{}												&
	\colhead{(K)}											&
	\colhead{(cm~s$^{-2}$)}									&
	\colhead{(Gyr)}											&
	\colhead{(M$_{\odot}$)}									&
	\colhead{(K)}											&
	\colhead{(cm~s$^{-2}$)}									&
	\colhead{(Gyr)}											&
	\colhead{(M$_{\odot}$)}                                
}
\startdata
2M1047				  	 	& T6.5 		& 888$^{+33}_{-33}$  		& 5.34$^{+0.11}_{-0.46}$ 	&  $>$2.5  				& $>$0.026					& 869$^{+35}_{-29}$	&  5.29$^{+0.10}_{-0.28}$    	& $>$2.5					& $>$0.026    				\\[6pt]
							&			& 850$^{+62}_{-47}$  		& 5.23$^{+0.18}_{-0.25}$ 	&  $>$2.5				& $>$0.026					&					&							&						&				  			\\[10pt]
SIMP0136					 	& T2.5 		& 1104$^{+51}_{-63}$ 		& 4.78$^{+0.35}_{-0.40}$ 	& 0.6$^{+1.1}_{-0.3}$    	& 0.022$^{+0.015}_{-0.012}$	& 1089$^{+62}_{-54}$	& 4.79$^{+0.26}_{-0.33}$  	& 0.6$^{+1.1}_{-0.3}$		& 0.022$^{+0.015}_{-0.012}$ 	\\[6pt]
							&			& 1073$^{+112}_{-87}$		& 4.79$^{+0.39}_{-0.52}$	& 0.7$^{+1.1}_{-0.3}$		& 0.022$^{+0.015}_{-0.012}$	& 					&							&						&							\\[10pt]	
2M1043\,\tablenotemark{d}  	& L8   		& 1012$^{+64}_{-90}$ 		& 3.94$^{+0.13}_{-0.09}$ 	& 0.6$^{+3.4}_{-0.3}$    	& 0.011$^{+0.011}_{-0.005}$	& 1390$\pm$180 		& \nodata                		& 0.6$^{+4.6}_{-0.3}$		& 0.011$^{+0.011}_{-0.005}$ 	\\[6pt]
							&			& 1229$^{+212}_{-260}$	& 4.28$^{+0.49}_{-0.34}$	& 0.6$^{+4.6}_{-0.3}$		& 0.011$^{+0.011}_{-0.005}$	&					&							&						&							\\[10pt]
2M1237						& T6.5		& 851$^{+36}_{-32}$		& 5.39$^{+0.08}_{-0.26}$ 	& $>$3.4					& $>$0.028                 	& 831$^{+31}_{-27}$	& 5.34$^{+0.08}_{-0.17}$  	& $>$3.4					& $>$0.028 					\\[6pt]
							&			& 810$^{+51}_{-43}$		& 5.28$^{+0.15}_{-0.21}$	& $>$3.4					& $>$0.028					&					&							&						&							\\[10pt]	
SDSS1254						& T2   		& 1079$^{+56}_{-63}$		& 4.52$^{+0.41}_{-0.35}$ 	& 0.49$^{+0.51}_{-0.21}$ 	& 0.017$^{+0.015}_{-0.008}$ 	& 1070$^{+69}_{-52}$	& 4.57$^{+0.30}_{-0.27}$		& 0.49$^{+0.51}_{-0.21}$ 	& 0.017$^{+0.015}_{-0.008}$	\\[6pt]
							&			& 1061$^{+127}_{-83}$		& 4.62$^{+0.43}_{-0.40}$ 	& 0.49$^{+0.48}_{-0.21}$	& 0.017$^{+0.015}_{-0.008}$	&					&							&						&							\\[10pt]	
SDSS0423\,\tablenotemark{e}	& L7+T2.5	& 1084$^{+71}_{-41}$ 		& 4.25$^{+0.34}_{-0.18}$ 	& 0.42$^{+0.62}_{-0.17}$ 	& 0.015$^{+0.021}_{-0.006}$ 	& 1678$^{+174}_{-137}$& \nodata 					& 0.43$^{+0.62}_{-0.17}$	& 0.015$^{+0.021}_{-0.006}$	\\[6pt]
							&			& 1150$^{+198}_{-114}$	& 4.50$^{+0.57}_{-0.35}$ 	& 0.43$^{+0.61}_{-0.17}$	& 0.014$^{+0.020}_{-0.006}$	&					&							&						&							\\[12pt]
Gl 570D					 	& T7.5 		& 817$^{+32}_{-36}$  		& 5.02$^{+0.19}_{-0.48}$  & 2.4$^{+1.6}_{-1.7}$    	& 0.024$^{+0.011}_{-0.010}$ 	& 799$^{+40}_{-32}$	& 4.96$^{+0.18}_{-0.32}$ 		& 2.4$^{+1.6}_{-1.7}$		& 0.024$^{+0.011}_{-0.010}$	\\[6pt]
							&			& 781$^{+73}_{-53}$		& 4.90$^{+0.32}_{-0.37}$	& 2.4$^{+1.6}_{-1.7}$		& 0.024$^{+0.011}_{-0.010}$	&					&							&						&							\\[12pt]
HN Peg B					 	& T2.5 		& 1054$^{+51}_{-66}$ 		& 4.60$^{+0.37}_{-0.44}$ 	& 0.6$^{+0.6}_{-0.3}$		& 0.018$^{+0.016}_{-0.009}$ 	& 1043$^{+59}_{-51}$	& 4.64$^{+0.28}_{-0.32}$ 		& 0.6$^{+0.6}_{-0.3}$		& 0.018$^{+0.017}_{-0.009}$ 	\\[6pt]
							&			& 1032$^{+107}_{-77}$		& 4.67$^{+0.40}_{-0.45}$ 	& 0.6$^{+0.6}_{-0.2}$ 	& 0.017$^{+0.015}_{-0.009}$	&					&							&						&					
\enddata
\tablenotetext{a}{(Top) cf. Gl~570D, \hspace{2mm}(Bottom) cf. HN~Peg~B.  Calibrators Gliese~570D and HN~Peg~B included for reference.  Minus and plus errors define the 68.27\% confidence interval.}
\tablenotetext{b}{Mass and age estimates from evolutionary models of \citet{baraffe2003}, using input parameters determined from (top) cf. Gl~570D and (bottom) cf. HN~Peg~B.  Minus and plus errors define the 68.27\% confidence interval, determined from 10,000 samples.  In cases where $>$20\% of input parameter samples fall outside of the \citet{baraffe2003} models, lower limits are within 84.13\% confidence.}
\tablenotetext{c}{Adopted values are averages from cf. Gl~570D and cf. HN~Peg~B, except for 2M1043 and SDSS0423.}
\tablenotetext{d}{Assuming no detection of Li in the optical spectrum in \citet{cruz2007}. Due to poor fit calibration for this object, we adopt instead T$_{\mathrm{eff}}$ calculated by applying the \cite{liu2010} bolometric correction to 2MASS H-band magnitude, typical brown dwarf radius $0.90 \pm 0.15 R_{\mathrm{J}}$, and conservative mass estimate $70 \pm 10$~M$_{\mathrm{J}}$.  We do not adopt a value for log~g and instead use the adopted mass and radius to calculate $<\rho>$ in Figure \ref{fig:christensenFig2}.}
\tablenotetext{e}{Parameter fits are based on the unresolved spectrum of the binary system and are thus highly suspect.  We adopt instead T$_{\mathrm{eff}}$ calculated from bolometric magnitude in \cite{vrba2004}, typical brown dwarf radius $0.90 \pm 0.15 R_{\mathrm{J}}$, and conservative mass estimate $70 \pm 10$~M$_{\mathrm{J}}$. We do not adopt a value for log~g and instead use the adopted mass and radius to calculate $<\rho>$ in Figure \ref{fig:christensenFig2}.}
\end{deluxetable*}

Effective temperatures (T$_{\mathrm{eff}}$) and surface gravities (log~g) were estimated for our sample following an updated version of the method described in \citep{burgasser2006b}. We used low-resolution near-infrared spectra from (a) the SpeX Prism Library \citep{burgasser2014_spex}; (b) data from \citealt{cruz2004, burgasser2004, liebert2007, siegler2007, burgasser2008}) and (c) the indices H$_2$O-J and $K/H$ defined in \citep{burgasser2006a, burgasser2006b}, which are orthogonally sensitive to temperature and surface gravity variations in T dwarf near-infrared spectra. The indices were measured on solar metallicity BTSettl08 spectral models \citep{allard2011} spanning T$_{\mathrm{eff}}$~=~600--1300~K and log~g~=~3.5--5.5~dex (cgs units). To calibrate these indices, we used the spectra of two brown dwarf companions with broad-band model-fit parameters: Gliese~570D (T7.5; \citealt{burgasser2000}) for which \citet{geballe2001} determine T$_{\mathrm{eff}}$~=~804$\pm$20~K and log~g~=~5.14$\pm$0.14~dex; and HN~Peg~B (T2.5; \citealt{luhman2007}) for which \citet{leggett2008} determine T$_{\mathrm{eff}}$~=~1115~K and log~g~=~4.81~dex. Scaling the corresponding model indices to be in agreement with these sources, we then identified the locus of model parameters for which these indices agree with the measured values for our six sources to within 3$\sigma$.  

Results are shown in Table~\ref{table:physparam}, which compares values from each of the calibrators separately.  For 2M1047, SIMP0136, 2M1237, and SDSS1254 we adopt the mean parameters from both Gliese~570D and HN~Peg~B calibrations. Note that values for 2M1237 are in agreement with those reported in \citet{liebert2007}, while we find a slightly cooler T$_{\mathrm{eff}}$ for SDSS1254 and a log~g on the low end of values reported by \citet{cushing2008}. The uncertainties for 2M1043 are fairly large and are most likely due to substantial differences between source and calibrator spectral types (a suitable late L dwarf calibrator was not available). Finally, while we report results for SDSS0423, these are highly suspect given the binary nature of this source \citep{burgasser2005b}.  Reported parameter uncertainties reflect uncertainties in the parameters selected to represent the calibrators Gliese~570D and HN~Peg~B and define the lower and upper bounds of the range relative to the central value that account for 68.27\% of the set.

The high surface gravities inferred for 2M1047 and 2M1237 indicate old ages and relatively high (substellar) masses. These were estimated from evolutionary models of \citet{baraffe2003} by drawing ten thousand T$_{\mathrm{eff}}$--log~g pairs from each distribution to determine the mean and standard deviations.  In both cases, $>$50\% of input parameter samples fall outside of the \citet{baraffe2003} models and may result in significantly skewed mean values, so we give lower limits within 84.13\% confidence.  For these sources we infer ages of $>$2.5 and $>$3.4~Gyr and masses of $>$0.026 and $>$0.028~M$_{\odot}$ within 84.13\% confidence, respectively.  In contrast, SDSS1254 is matched to a very young age ($\sim$500~Myr) and low mass $\sim$0.017~M$_{\odot}$). Note that \citet{cushing2008} report disagreement in log~g values based on evolutionary models (log~g~=~4.7--4.9) and spectral model fits (log~g~=~5.0--5.5), which these authors speculate may be due to unresolved multiplicity. Our difficulties in inferring the properties of 2M1043 may be related to this source's unusual cloud properties, as it is one of the reddest L8 dwarfs known ($J-K_s=1.97\pm$0.08). Its reported optical spectrum shows no indication of Li~I absorption \citep{cruz2007} implying a mass $\sim$0.011~M$_{\odot}$ and age $\sim$600~Myr, although this feature may have been masked by poor continuum detection. 

For objects whose parameters are not well constrained by the above method, we follow \cite{vrba2004} and adopt a typical radius of $0.90 \pm 0.15~R_{\mathrm{J}}$ from the \cite{burgasser2001} study of radius distribution in \cite{burrows1997} L and T dwarf evolutionary models.  We adopt a typical late-L mass range of $70 \pm 10$~M$_{\mathrm{J}}$.  For 2M1043, we apply a bolometric correction calculated for spectral type L8 using the polynomial fit from \cite{liu2010} to the 2MASS H-band magnitude.  Using $M_{\odot,\mathrm{bol}}=4.7554 \pm 0.0004$~mag and $L_{\odot,\mathrm{bol}}\,\footnote{Adopted from Eric Mamajek's Star Notes: \\ https://sites.google.com/site/mamajeksstarnotes/basic-astronomical-data-for-the-sun}=3.827 (\pm0.0014)\times10^{33}$~erg~s$^{-1}$, we convert the bolometric magnitude to an effective temperature T$_{\mathrm{eff}} = 1390 \pm 180$~K.    For SDSS0423, we adopt T$_{\mathrm{eff}}=1678^{+174}_{-137}$~K as derived by \cite{vrba2004}.  We include these parameters in Table~\ref{table:physparam}.

\subsection{A Simple Formalism for Comparing Magnetic Field Measurements}\label{sec.formalism}
\subsubsection{Magnetic Field Topology}\label{sec.dipole}
Radio observations of highly circularly polarized pulsed emission yield precise measurements of local magnetic field strengths in the magnetospheres of our objects.  However, translating them to a global field strength useful for evaluating dynamo models requires topological information that is difficult to determine from radio observations alone.  

\citet{lynch2015} attempted to constrain the field topologies for two pulsing radio dwarfs by modeling their radio dynamic spectra, inferring localized loops and loss-cone ECM from their modeling.  In contrast, \citet{kuznetsov2012} similarly model the radio pulses of one of the dwarfs examined by \citet{lynch2015} and found that a highly inclined dipole model with active longitudes for shell-type electron distributions reproduces the pulses with greater fidelity than a loss-cone distribution.  Others have inferred dipole-dominated \citep{yu2011}, quadrupole-dominated \citep{berger2009}, or small-scale-dominated \citep{cook2014, williams2014} field geometries for pulsing radio dwarfs.  Similar extrapolations have been made for Jovian radio emission using ExPRES (Exoplanetary and Planetary Radio Emissions Simulator) by \cite{hess2008, hess2011}. However, the latter use a plethora of additional information to help constrain their calculation, including information on the radio source distribution, the beaming in the planetary environment, a planetary magnetic field model, and precise knowledge of the planetary inclination to the line of sight, none of which are currently available for the dynamic spectra of ultracool dwarfs.  We do not attempt to recover the field topologies of our objects here. 

Instead, we consider the case where a dipole drives the observed emission.  Although direct confirmation of the electrodynamic engine(s) at work in our objects is required to infer whether our magnetic field measurements are indeed of the dipole component or are instead from higher order components, we note that detailed observations of the magnetized Solar System planets show that the dipole component is most likely to produce auroral emission.  Specifically, interactions between the large-scale planetary magnetic field with the solar wind \citep{isbell1984}, the planetary field with orbiting moons such as the Jupiter-Io current system \citep{goldreich1969}, and co-rotation breakdown of a plasma sheet in the planetary magnetosphere drive the electrodynamic engines of the Solar System planets \citep[and references therein]{hill2001, cowleyBunce2001, bagenal2014, badman2015}.  In all cases, energy is coupled into the upper atmosphere from distances sufficient for the planetary dipole components to dominate. 

For our objects, magnetosphere-ionosphere (M-I) coupling via co-rotation breakdown or satellite interaction have been proposed as likely drivers \citep{schrijver2011,nichols2012, hallinan2015}.  We first consider satellite interaction.  For a brown dwarf with a rocky satellite, the Roche limit occurs at $\sim$3.7R$_*$ \citep{murrayDermott1999}.  Even at this minimum distance, dipole fields dominate over higher order fields that are a factor of 3 stronger at the surface.  In comparison, corotation breakdown occurs at 30--50R$_{\mathrm{J}}$ for Jupiter \citep[and references therein]{cowleyBunce2001, hill2001, vogt2011}, and at 3--4R$_{\mathrm{S}}$ for Saturn \citep{stallard2010}.  In these cases, dipole fields of surface field strengths $\sim$2--50 times weaker than a quadrupole surface field would dominate at the corotation breakdown radius.

Zeeman Doppler imaging by \citet[hereafter JM10]{morin2010} suggests that objects significantly below the fully convective boundary with $\sim$kilogauss large-scale fields are dipole-dominated, with the majority of their magnetic energy lying in the dipole component.  Specifically, they find that magnetic topologies of 11 M5--M8 dwarfs fall into either a strong or weak large-scale field regime (strong LSF and weak LSF, respectively).  In the strong LSF regime, the large-scale field is of order kilogauss with 66--90$\%$ of the reconstructed magnetic energy in the dipole component and is temporally stable over at least $\sim$3 years, the length of the study.  In the weak LSF regime, multipolar field topologies with much weaker $\sim$0.1~kG large-scale fields vary significantly on year timescales.  If the results of \citet{morin2010} apply to late L and T dwarfs, then objects in the strong LSF regime are unlikely to host quadrupolar fields a factor of three or more times stronger than the dipole component, and the dipole field would drive the M-I coupling currents.  

In contrast, \citet{williams2014} argue that weak LSF objects may be X-ray dim/radio bright (departing from the G{\"u}del-Benz relation) instead of X-ray bright/radio dim (more aligned to the G{\"u}del-Benz relation).  They suggest that objects in the weak LSF regime likely experience less magnetic activity than objects in the strong LSF regime, hypothesizing that the decreased magnetic activity in weak LSF objects result in correspondingly underluminous X-ray emission, but that small-scale reconnection events can provide a source of radio-emitting electrons.  However, we note that in the standard reconnection model of chromospheric heating, X-ray and radio luminosities are tightly correlated \citep[and references therein]{gudelBenz1993, gudel2002, benz2010, forbrich2011}, except for extremely small solar flares, which are in fact comparatively radio underluminous rather than X-ray dim/radio bright.  Accordingly, the presence of small-scale reconnection events from a strong small-scale field (as in the weak LSF regime) would result in objects that adhere more closely to the G{\"u}del-Benz relation.  

Instead, the lowering of fractional ionization can explain the relative decrease in X-ray luminosities \citep{mohanty2002}.  This does not necessarily impact the radio emission, which is produced above the photosphere or chromosphere irrespective of the mechanism by which it is produced and does not necessarily have the same dependence on fractional ionization as coronal heating.  It is also important to note that previous Zeeman broadening studies for 9 of the 11 stars studied in JM10 measured mean surface field magnitudes of order kilogauss \citep{reinersBasri2007, reinersBasriBrowning2009}, regardless of which field regime the star occupied.  This implied that the small-scale fields rather than the large-scale ones are quite strong in the weak-field regime. However, in such a scenario, we note that even though the current understanding of M-I coupling does not require the fields to be dipolar, they must be large-scale and strong (kilogauss or stronger to fit observations), precluding the possibility that even strong small-scale fields could drive the M-I coupling. 

In the case that JM10 does not extend to our objects, late L and T dwarfs may in fact be more analogous to gas giant planets than to M-dwarfs.  Jupiter and Saturn are both dipole-dominated, with the quadrupole and octupole moments at $\sim$20\% of the dipole moment in Jupiter \citep{acunaNess1976}, and the quadrupole moment in Saturn only $\sim$7\% of its dipole moment \citep{russell1993}.  Despite significant higher order moments present in the Jovian field, the auroral radio emission produced by Jupiter is thought to be dominated by the dipolar field component \citep{hill2001}. 

While it is possible for higher order components to drive M-I coupling currents, it is clear that the dipole field can efficiently generate auroral currents. Therefore, we treat the dipole case and will revisit alternatives when additional information on the magnetic fields of ultracool dwarfs becomes apparent.

\subsubsection{Relating Magnetic Fields Measured from Auroral Radio Emission to Zeeman Techniques }

Under the assumption that auroral emission can be associated with the dipole component of the magnetic field, we now relate our magnetic field measurements to those obtained from Zeeman broadening and Zeeman Doppler imaging observations so that we may compare our ECM measurements to the \citet{christensen2009} results, which use Zeeman-based measurements.  To begin, it is important to understand what information each technique yields and its limitations, and we refer the reader to more detailed discussion in \cite{reiners2012} and \cite{morin2012} and the references therein. 

Zeeman broadening measurements from spectral observations of magnetically sensitive lines provide mean surface field magnitudes $B_{\mathrm{s}}$, averaged over the photospheric surface of stars, or in rare cases, averaged over the magnetically active regions of the star.  For stars where the Zeeman splitting of the $\sigma$ components can be resolved, both the mean magnetic field magnitude $B_{\mathrm{s}}$ and filling factor $f$ may be measured from the magnitude of the splitting and the relative depths of the $\sigma$ and $\pi$ components, respectively \citep{valenti1995,johnskrullValenti1996, johnskrull2000ASPC}.  This requires atomic lines to be relatively isolated for comparison with continuum flux.  M5 or later type objects suffer from spectra increasingly contaminated by molecular lines, and lines become dominated by pressure broadening. In cases where the Zeeman splitting cannot be resolved from the intrinsic line width, the filling factor remains entangled with the mean field and it is possible to measure only $B_{\mathrm{s}} = B_{Z}f$. \cite{reinersBasri2007} were able to measure mean field magnitudes by comparing the FeH features of 24 M2--M9 stars to reference spectra with known $B_{Z}f$, with $\sim$15\%--30\% uncertainties \citep{reiners2012, shulyak2010}.  The method described by \cite{reinersBasri2006} is limited by the reference spectra; $B_{Z}f$ is measured in reference to a zero field spectrum and a 3.9~kG spectrum, so only fields less than 3.9~kG can be quantified, though it is unlikely that the object serving as the zero field reference is in fact magnetically inactive. Finally, Zeeman broadening techniques have yet to be successfully applied to objects beyond M9, where rotational broadening blends useful molecular lines.  Despite limitations, Zeeman broadening provides a straightforward and convenient framework within which to interpret measurements when testing dynamo predictions. 

Zeeman Doppler imaging (ZDI) provides approximate reconstructions of surface field topologies, allowing estimates of the magnetic energy in different field components (for example, the dipole). However, as applied with existing instruments, ZDI measurements are only sensitive to larger-scale fields, especially in very dim and fast rotators such as our objects.  The sensitivity of ZDI is limited by current abilities to adequately resolve polarized flux.  Inadequate resolution can lead to the apparent canceling out of observational signatures of opposite polarity fluxes and mask magnetic fields at smaller spatial scales.  For this reason, ZDI is more sensitive to large-scale field structures that can be fully resolved by existing instruments \citep{reinersBasri2009, yadav2015}, and JM10 have found that the dipole energy can vary by $\sim$10--30\%, with significant confusion between the dipole and quadrupole components.  Additionally, instruments used to map the magnetic fields of cool stars were limited to 2 of the 4 Stokes parameters (I, V) until very recently \citep{rosen2015}, which further limits the sensitivity of ZDI in fully capturing magnetic field topologies.  Finally, ZDI maps can vary widely depending on the particular entropy weighting prescription used when phase coverage is insufficient.  Nonetheless, the sensitivity of ZDI to large-scale fields has provided vital insight into large-scale fields.  Field topologies of stars appear to change from being dominated by a weak non-axisymmetric toroidal field to a strong axisymmetric poloidal field as they cross into the fully convective regime \citep{donati2008, morin2008}, and JM10 found evidence for bistable field topologies in late-M dwarfs, as discussed in \S \ref{sec.dipole}.  

Using either of the Zeeman techniques to measure magnetic fields is currently impossible for objects beyond spectral type M9, yet the mass regime occupied by L and T dwarfs is critical for probing the efficacy of any fully convective dynamo model. Radio observations of ECM emission provide a new window for probing magnetic activity in a mass regime where Zeeman broadening techniques cannot currently reach.  Because the measured magnetic field magnitudes are dependent only on the frequency of the emission cutoff, measurements from radio observations are not subject to the same sources of uncertainty that affect the accuracy of Zeeman broadening measurements. However, ECM measurements also have limitations. Rather than measuring an average field strength, radio observations give a single measurement with great accuracy of the local magnetic field strength in the region of the magnetosphere corresponding to the emission.  Additionally, they are likely primarily sensitive to large-scale fields and the data in isolation are not sufficient for reconstructing the field topology.  Finally, without observing emission cutoffs, we are limited to interpreting our measurements as lower-bounds to global maximum surface field strengths. 

To estimate the lowest possible bound on the global rms surface field strength of an object from a single local radio-derived measurement, we consider an idealized dipole case, which we will adjust as additional topological information becomes available. Our interest in obtaining a conservative lower limit allows us to assume the following simplifications for all of our objects: 
\begin{enumerate}
\item The magnetic field is perfectly dipolar (the presence of higher order fields will positively contribute to the rms surface field).
\item The lower bound field strength measured from our ECM observations, B$_{\mathrm{ECM}}$, is the field strength at the magnetic pole at the photosphere.  In reality, the emission likely samples the field at a location that does not correspond exactly with the magnetic pole.  Moreover, until we observe a frequency cutoff, the emission corresponds to a location in the magnetosphere that is a nonzero altitude above the photosphere, so the actual surface polar field strength can only be equal or greater in all cases.
\item Brown dwarfs are perfect spheres.
\end{enumerate}
We calculate the mean surface dipole field, beginning with the expression for a dipole field, 
\begin{equation} 
\vec{B}(\vec{r}) = \frac{\mu_0}{4\pi} \left[ \frac{3 \hat{n} (\hat{n} \cdot \vec{m}) - \vec{m}}{|\vec{r}|^3} \right]
\end{equation}
where $\hat{n} = \vec{r}/|\vec{r}|$ is the unit vector in the direction to the point on the sphere for which the field strength is calculated and $\vec{m}$ is the magnetic dipole moment.  Averaging over the surface of the star shows that the mean squared surface field strength due to the dipole field is
\begin{equation}
\langle B_{\mathrm{s, dip}}^2 \rangle = \frac{1}{2} B^2_{\mathrm{ECM}} \;.
\end{equation} 

In the case where our objects have purely dipolar fields, $ \langle B_{\mathrm{s, dip}} \rangle ^{1/2}$ would be equivalent to the mean surface field magnitude $B_{\mathrm{s}} = B_{Z}f$ as measured by Zeeman broadening, with a filling factor of 100\%.   Where our objects do not have purely dipolar fields, we consider two cases.  If higher order fields are anti-aligned with the dipole field, such that they contribute negatively to the magnetic flux at the pole, then $\langle B_{\mathrm{s, dip}} \rangle^{1/2}$ as calculated above will underestimate the lower bound of the mean surface field magnitude.  If higher order fields are aligned with the dipole such that they contribute positively to the flux at the magnetic pole, then the field strength measured from radio emission will overestimate the rms surface dipole field. 

To understand the severity of such a possible overestimation, we return to the \cite{morin2010} study.  While \cite{morin2011, morin2011sf2a} interpret the result as possible evidence for a dynamo bistability, \cite{kitchatinov2014} have also proposed that it is evidence of an M-dwarf magnetic cycle.  No objects have been observed to be in a transition between the strong field and weak field regimes, suggesting that if such a transition occurs, as in a magnetic cycle, the transition is very fast and is unlikely to impact the interpretation of our field measurements.  We know from the observed ECM emission and our discussion in \S \ref{sec.dipole} that our objects likely occupy the strong LSF regime of a possible bistable dynamo or magnetic cycle.  This implies relatively weak higher order fields, limiting any overestimation of the mean surface field magnitude.

\subsection{An Application to Dynamo Models: \\ Comparison to Christensen 2009 Model}
\begin{figure*}
\epsscale{.8}
\plotone{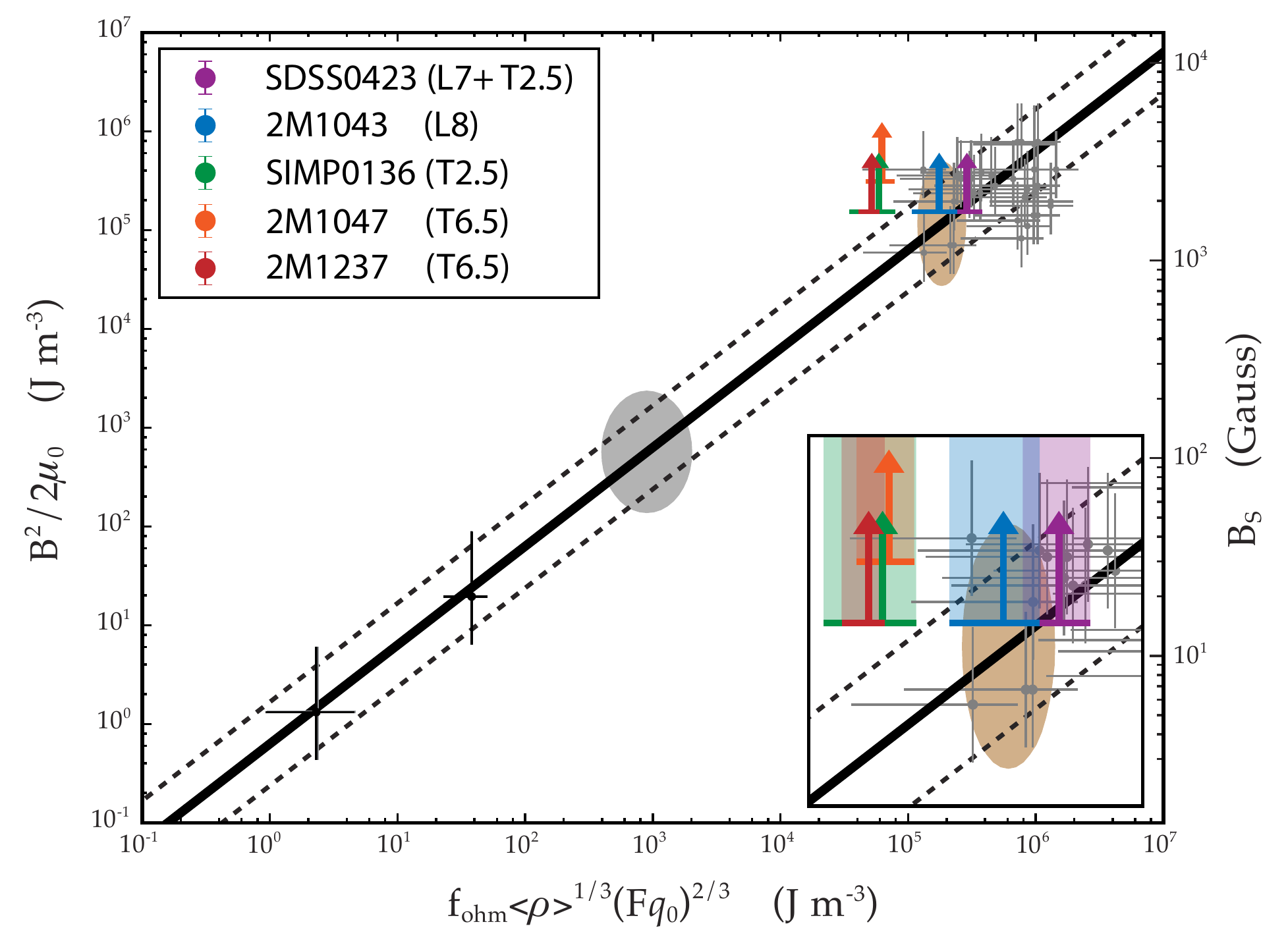}
\caption{\label{fig:christensenFig2}
Reproduction of Figure 2 from \cite{christensen2009}, showing their proposed dynamo scaling relation with 3$\sigma$ uncertainties for fully convective, rapidly rotating objects (black solid line and dashed lines, respectively). Grey points represent TTauri stars and old M-dwarfs.  Black points represent Earth and Jupiter.  The brown ellipse indicates the predicted position for a 1500~K brown dwarf and  the grey ellipse indicates the predicted position for a 7 M$_{\mathrm{J}}$ exoplanet.   Our detected targets are overplotted, with upward arrows to indicate that our measurements are lower bounds and horizontal bars to indicate estimated uncertainties. The inset shows more clearly our estimated uncertainties.  We adopt a minimum surface field strength of 2.5~kG for our newly detected objects.  For 2M1047, we adopt 3.6~kG as measured by \cite{williams2015}.}
\end{figure*}

We now attempt to test the scaling law presented by \cite{christensen2009} (hereafter C09).  C09 showed that for planets and fully convective and rapidly rotating (P$<$4 days) stars, the convected energy flux available may generate the magnetic field strength.  In a departure from prevailing dynamo scaling laws, the central tenet to their model was an energy balance between kinetic and magnetic energies and ohmic dissipation and convective heat transport, rather than a force balance between the Coriolis, Lorentz, buoyancy, and pressure forces \citep{christensenAubert2006}.  Surprisingly, they found that the magnetic field strength is independent of both magnetic diffusivity and rotation rate and instead depends strongly only on the buoyancy flux and dynamo size.  In particular, they show that for Jupiter, Earth, and a sample of stars including T Tauri stars, old M-dwarfs, and main sequence stars with P$<$4 days, the following relation is empirically consistent: 
\begin{equation}
\langle B^2\rangle/(2\mu_0) = c f_{\mathrm{ohm}} \langle\rho\rangle^{1/3} (Fq_0)^{2/3} \; .
\end{equation} 
Here, $\langle B^2\rangle$ is the squared magnetic field averaged over the whole volume of the dynamo region rather than the surface of the star. $f_{\mathrm{ohm}}$ is the ratio of ohmic dissipation to total dissipation and is nominally assumed to be $f_{\mathrm{ohm}} \approx 1$.  $F$ is a volume average of the temperature scale height divided by the length scale of the largest convective structures, and for their purposes, C09 assume $F=1$ and 1.19 for stars and Jupiter, respectively.  For the purposes of our analysis, we adopt $F=1$.  $q_0$ is the bolometric flux at the outer boundary of the dynamo regions, which C09 obtain from the effective surface temperatures of the stars. Finally, $\mu_0$  is permeability,  $\langle\rho\rangle$ is the mean density of the dynamo region, and $c$ is a proportionality constant.  Figure \ref{fig:christensenFig2} reproduces this scaling law.  Significantly, the wide mass range that the above empirical relation describes tantalizingly hints that the scaling law may be generalizable for all convection-driven dynamos. 

The C09 model calls for the mean internal field strength $\langle B \rangle$ of dynamo regions, and an ideal test of their model would utilize direct measurements of the magnetic field inside of the dynamo itself.  However, measuring these data is impossible.  Instead, they estimate $\langle B \rangle$ in several ways.  The most direct observational tests available to C09 are Zeeman broadening measurements from spectral observations of Ti~I lines in T Tauri stars by \cite{johnskrull2007} and K and M stars by \cite{saar1996} and FeH lines in M-dwarfs by \cite{reinersBasri2007}.  C09 additionally adapt ZDI data of mid M-dwarfs by \cite{morin2008}.  

The lower bound mean surface field magnitude $B_{\mathrm{s, dip}}$ that we calculated for our objects allows us to very straightforwardly compare our field measurements with those predicted by C09.  We treat $B_{\mathrm{s, dip}}$ for each object as a lower bound Zeeman broadening measurement $B_{\mathrm{s}}$ and convert it to $\langle B \rangle$ by following C09 and multiplying by a factor of $\langle B \rangle/B_{\mathrm{s}}\approx 3.5$, which they report is the typical ratio found in their geodynamo simulations. In a recent study of 2M1047, \cite{williams2015} detected a pulse at $\sim$10~GHz, corresponding to a lower bound surface field strength of 3.6~kG for this object. We adopt this value in our comparison to field strengths predicted by C09. 

We overlay our most conservative field constraints from auroral radio emission on our reproduction of the C09 scaling law in Figure \ref{fig:christensenFig2}. All of our T dwarfs depart mildly from the C09 scaling relation, suggesting four possibilities: (1) parameters beyond convective flux and dynamo size may influence magnetic fields in brown dwarfs, (2) brown dwarfs have a systematically larger value for the parameter converting external field to internal field, (3) their fields are systematically stronger at the poles than what a dipole predicts, or (4) their field topologies are not dominated by dipoles. These possibilities would not necessarily undermine the basic premises of the proposed scaling law but simply add more uncertainty to the precision with which it can applied. 

It is important to remember that dynamo scaling laws are powerful tools for elucidating which general physical characteristics and behaviors matter, but they describe an inherently chaotic process and the laws are not deterministic. It is possible that C09 may in fact be largely conceptually correct in the scaling law that they propose, but the parameters on which their law depends may differ from group to group.  For instance, the dynamo region extends over $\sim$6--10 orders of magnitude in density in low mass stars \citep{saumon1995}.  The outermost part of the dynamo action is in a region that is much less dense than the mean density of the dynamo region, yet that could well be the region that determines the observed field because it is closest to the outer boundary.  Another possibility is that the appropriate density to use may be defined differently between brown dwarfs and low mass stars. Additionally, parameters such $\langle B \rangle/B_{\mathrm{s}}$ depend on boundary conditions, rotation rate, density structure, specific properties of the outer insulating shell (present in Jupiter and brown dwarfs, but not in low mass stars), etc.  Finally, the C09 model is specific to dipole-dominated fields ($>$35\% of field strength in the dipole component), so a departure from the relation may indicate field topologies dominated by higher-order fields. 

Nevertheless, it is notable that some of our objects have lower bound field strengths that are systematically higher than what C09 predict when using parameter definitions that they adopted. The dynamo surface in Jupiter is at $\sim$0.85$R_{\mathrm{J}}$ \citep{guillot2004}, whereas it is near the surface of M-dwarfs.  For our objects, the dynamo surface may be more interior than in M-dwarfs, causing the adopted values of $q_0$, $\langle\rho\rangle$, and $B_s$ to increase.  However, $B^2$ rises faster than $\langle\rho\rangle^{1/3}(q_0)^{2/3}$ as a function of internal radius, independent of field topologies, so our T dwarfs may in fact depart more dramatically. Pushing subsequent studies to higher frequencies to observe emission cutoffs will be necessary to obtain the best possible constraints for field measurements derived from auroral radio emission.

\subsection{Implications of Auroral Radio Emission Correlated with Brown Dwarf Weather and H$\alpha$ Emission}
Prior to our work, radio surveys of $\sim$60 $\geq$L6 objects yielded only one detection \citep{antonova2013, route2013}, resulting in a detection rate of just $\sim$1.4\%.  In contrast, we have achieved a notably higher detection rate of 4/5 objects, not including the previously-detected 2M1047, by departing from previous target selection strategies and biasing our targets for previously confirmed H$\alpha$ emission, or in the case of SIMP0136, optical/IR variability.  Several of our objects also exhibited tentative IR variability.  Selection effects from inclination angles or increased instrument sensitivity may contribute to our dramatically higher success rate, but it is also clear that biasing our sample for optical auroral emission provides a good means to finding radio-emitting brown dwarfs.  

While the relationship between IR variability and auroral radio emission remains uncertain, our results are intriguing when viewed in the context of brown dwarf weather.  J-band variability appears to be common in L and T dwarfs \citep{enoch2003, clarke2008, radigan2014a, radigan2014b, buenzli2014, metchev2015}. Included in our target sample is the canonical dust-variable T-dwarf SIMP0136, which exhibits large-amplitude ($>$5\%) IR variability. Also included were tentatively low-amplitude variable objects SDSS0423, 2M1237, and SDSS1254.  Clouds in brown dwarf atmospheres have been proposed to interpret observed photometric and spectroscopic variability, and where objects have been observed at multiple wavelengths, some proposed models rely on patchy clouds of variable thicknesses and temperatures \citep{marley2010, burgasser2014_luhman16ab, apai2013} to explain wavelength-dependent variability.  Our results point to the possibility that an additional variability mechanism may be at play, as postulated by \cite{hallinan2015}.

The success of our selection strategy is especially compelling in light of simultaneous radio and optical spectroscopic observations of the M8.5 dwarf LSR~J1835+3259 (hereafter LSR~J1835) by \cite{hallinan2015}, whose results in fact motivated our selection strategy.  Their study shows features in the radio dynamic spectrum and in the optical spectrum that vary either in phase or anti-phase with each other, with a 2.84-hr period that corresponds to the known rotation period of LSR~J1835.  \cite{hallinan2015} assert that auroral current systems can explain the Balmer line emission and observed multi-wavelength periodicity.  Specifically, they argue that the downward spiraling population of electrons that gives rise to the observed ECM emission also causes collisional excitation of the neutral hydrogen in the atmosphere upon impact, with subsequent de-excitation via line emission powering the observed Balmer emission.  Additionally, the electron current supplies the brown dwarf atmosphere with excess free electrons, possibly contributing to increased H$^-$ opacity in the auroral feature.  The increased H$^-$ opacity would cause the upper atmosphere of the auroral feature to become optically thick, appearing lower in temperature than the photosphere. Such an auroral H$^-$ `cloud' could explain the phased and anti-phased lightcurves at various wavelengths observed in both LSR~J1835 and TVLM~513-46546 \citep{littlefair2008}, another M8.5 brown dwarf known to emit both quiescent and periodically pulsing radio emission as well as H$\alpha$, with a lasting $\sim$0.4-period offset between the optical emission and the radio pulses \citep{hallinan2007, berger2008, wolszczan2014, lynch2015}. 

Our results corroborate the unified auroral model proposed by \cite{hallinan2015} for even the coolest dwarfs.  In late-L and T dwarfs such as our targets, molecular hydrogen dominates the atomic hydrogen in the atmosphere, and observed photometric variability may in part be explained by localized heating of the atmosphere within the auroral feature by the precipitating electron beam.  \cite{morley2014} showed that heating of the atmosphere at different depths perturbs the pressure vs. temperature profile and can indeed cause spectral variability. Regardless of where in the atmosphere heating occurs, the highest amplitude variability occurs in absorption features redward of $\sim$2.2~$\mu$m, which could lead to variability in the K and L bands.  Encouragingly, K$_s$-band variability has been observed in SIMP0136, as well as tentatively for SDSS0423, and \cite{metchev2015} report that $36^{+26}_{-17}\%$ of T dwarfs vary by $\geq$0.4\% at 3--5~$\mu$m.  However, the incidence rate for dust variability is much higher than for auroral emission \citep{buenzli2014, radigan2014a, radigan2014b, metchev2015, heinze2015, kirkpatrick2000, burgasser2003_redOpticalData, cruz2007, kirkpatrick2008, pinedaInPrep}, suggesting that auroral emission may only play a role in some cases, such as the highly variable SIMP0136.   Finally, we note that even in the absence of atomic hydrogen, H$\alpha$ emission can still occur.  The incoming populations of free electrons and protons can recombine to excited states or the molecular hydrogen may dissociate to excited atomic hydrogen, subsequently de-exciting via Balmer emission. 

In addition to the possible correlation with IR variability, all previous detections of pulsed radio emission from ultracool dwarfs have been accompanied by detectable levels of quiescent radio emission, with no reported detections of pulsed emission in isolation. Although the properties of the quiescent emission are consistent with incoherent synchrotron or gyrosynchrotron emission, the physical processes governing the pulsed and quiescent emission are likely causally related, with the possibility of a shared electrodynamic engine powering the emission.

To better understand the relationship between H$\alpha$, radio, and IR variability, additional simultaneous multi-wavelength observations and detailed models investigating atmospheric heating from the auroral currents are needed.

\section{Conclusions}\label{sec.Conclusion}

We detected 5 of 6 late-L/T dwarfs in the 4--8 GHz band, including first detections for 4 objects, quintupling the number of radio-detected objects later than spectral type L6.  For 4 of our objects, including previously-detected 2M1047, we observe highly circularly polarized pulsed emission.  We also tentatively observe circularly polarized pulsed emission from a fifth object, 2M1237.  All of our objects with pulsed emission also exhibit quiescent emission, as is the case for all previously detected radio brown dwarfs.  This suggests that pulsed and quiescent phenomena are almost certainly related, though the mechanism for quiescent emission is still unclear.

Biasing our sample for H$\alpha$ emission or optical/IR variability provides a good means to finding these objects, implying that the H$\alpha$ emission may be the optical counterpart of auroral activity observed in the radio. We additionally note that several of our objects are either confirmed or tentative IR-variable sources, including the well-known dust variable SIMP0136.  Viewed in light of recent studies by \cite{hallinan2015} and \cite{morley2014}, our radio detections hint that auroral activity may also be related to brown dwarf weather in some cases.

Our data confirm kilogauss magnetic fields down to spectral type T6.5, demonstrating the efficacy of ECM as a tool for probing the magnetic fields of the coolest dwarfs in a mass gap that is critical for informing fully convective dynamo models. 

We develop a framework for comparing magnetic field measurements derived from electron cyclotron maser emission to measurements derived from Zeeman broadening and Zeeman Doppler imaging techniques. Using our framework, we provide strong constraints for rms surface field strengths in late-L/T dwarfs and demonstrate that our T dwarfs have magnetic fields that may be inconsistent with the \cite{christensen2009} model.  This suggests that parameters beyond convective flux may influence magnetic field generation in brown dwarfs.



\acknowledgments
\section{Acknowledgements}\label{sec.Acknowledgements}
MMK thanks Jackie Villadsen, E. Sterl Phinney, Ulrich Christensen, and Shri Kulkarni for illuminating and helpful conversations.

This publication makes use of data products from the Two Micron All Sky Survey, which is a joint project of the University of Massachusetts and the Infrared Processing and Analysis Center/California Institute of Technology, funded by the National Aeronautics and Space Administration and the National Science Foundation.

The National Radio Astronomy Observatory is a facility of the National Science Foundation operated under cooperative agreement by Associated Universities, Inc.

JSP was supported by a grant from the National Science Foundation Graduate Research Fellowship under grant no. DGE-1144469.



Facilities: \facility{JVLA}



\bibliography{bridgegap}


\clearpage

\end{document}